\documentclass[12pt, preprint]{aastex}
\usepackage{natbib}
\usepackage{graphicx}
\usepackage{wrapfig}
\usepackage[textheight=9in,textwidth=7in,headsep=0.5in,nofoot]{geometry}

\shortauthors{Kowalski et al.}
\shorttitle{SDSS M Dwarf Flare Rates}
\begin{document}

\title{M Dwarfs in SDSS Stripe 82:  Photometric Light Curves and Flare Rate Analysis$^1$ \footnotetext[1]{Based on observations obtained with the Apache Point Observatory 3.5-meter telescope, which is owned and operated by the Astrophysical Research Consortium.}
}

\author{Adam F. Kowalski\altaffilmark{2},
	 Suzanne L. Hawley\altaffilmark{2},
	 Eric J. Hilton\altaffilmark{2},
	 Andrew C. Becker\altaffilmark{2},
	 Andrew A. West\altaffilmark{3},
	 John J. Bochanski\altaffilmark{3},
         Branimir Sesar\altaffilmark{2}
       }

\altaffiltext{2}{Astronomy Department, University of Washington,
   Box 351580, Seattle, WA  98195\\
email: kowalski@astro.washington.edu}
\altaffiltext{3}{MIT Kavli Institute for Astrophysics \&
Space Research, 77 Massachusetts Ave, Building 37, Cambridge, MA
02139}

\begin{abstract}
  We present a flare rate analysis of 50,130 M dwarf light curves in SDSS Stripe 82. We identified 
271 flares using 
a customized variability index to search $\sim$2.5 million  photometric observations for flux increases in the $u$- and $g$-bands.  Every image of a flaring observation 
was examined by eye and with a PSF-matching and image subtraction tool to guard against false positives.  
Flaring is found to be strongly correlated with the appearance of H$\alpha$ in emission in the quiet spectrum.
Of the 99 flare stars that have spectra, we classify 8 as relatively inactive.  The flaring fraction is found to increase strongly 
in stars with redder colors during quiescence, which can be attributed to the increasing flare visibility and increasing active fraction for 
redder stars.  The flaring
fraction is strongly correlated with $|Z|$ distance such that most stars that flare are within 300 pc of the Galactic plane.  We derive 
flare $u$-band luminosities and find that the most luminous flares occur on the earlier-type M dwarfs.  Our
best estimate of the lower limit on the flaring rate (averaged over Stripe 82) for flares with $\Delta u$ $\ge$ 0.7 magnitudes on
stars with $u < 22$ is 1.3  flares hour$^{-1}$ square degree$^{-1}$  but can vary significantly with the line-of-sight.

\end{abstract}


\section{Introduction}\label{sec:intro}

Flares are a powerful manifestation of stellar magnetic fields,
releasing stored energy from surface magnetic loops
into the outer atmosphere of the star, and causing strong transient
increases in the blue and ultraviolet continuum emission as well as in 
optical, UV and X-ray emission lines.  There are many unsolved problems 
in flare physics, ranging from the details of the energy release, to
the mechanisms for producing the atmospheric emission, to the understanding
of flares on a global scale -- which stars flare, how often, how much
emission do they produce, and on what timescales?  In the past, the
investigation of these global questions has been hampered by the
need to obtain long time series observations of individual stars,
which is an exceedingly time-consuming exercise involving both ground-based data
\citep{Moffett1974, LME1976} and satellite data \citep{Audard2000, Gudel2003}.  The advent of large
time-resolved sky surveys covering a significant fraction of the sky (e.g. Pan-STARRS, \cite{Kaiser2004};
LSST, \cite{Tyson2002}, \cite{Ivezic2008LSST}; Gaia, \cite{Perryman2001})
will provide an entirely new opportunity to assemble flare
data from repeat observations of millions of stars on timescales of minutes to days.
The analysis of these new data will be quite different from classical
flare data which were obtained continuously on single objects.  


In this paper we use data from the Sloan Digital Sky Survey 
\citep[SDSS; ][]{York2000} to perform an initial examination of the
possibilities for studying flares in a large sky survey database.
The SDSS contains repeat observations of
one particular part of the sky, known as Stripe 82, along the Celestial
Equator.  This region was targeted because
it is in the South Galactic Hemisphere, and is visible from Apache Point
Observatory during the northern 
autumn months (Sep-Dec) when the main SDSS fields in the North Galactic 
Hemisphere are not accessible.  The original motivation was to obtain numerous 
photometric scans of
the same region of sky so that they could be co-added to produce a photometric
template 1-2 magnitudes deeper than the normal SDSS data and to quantify observational
precision and repeatability.  
It was recognized early on, however, that the individual observations formed 
a valuable time sequence of data that could be mined to identify time-variable 
phenomena.  Many additional SDSS-II scans of this region
were also obtained as part of the SDSS-II Supernova Survey 
\citep{Frieman2007}.  The Stripe 82 data products are contained in
the SDSS Data Release 7 \citep{Abazajian2009}.
An investigation aimed at characterizing flare signatures in the SDSS 
Stripe 82 data is therefore timely.  

We have two primary goals in this paper.
The first is to investigate flare frequencies as a function of spectral type
and magnetic activity status, where active stars are defined as those
with the chromospheric H$\alpha$ line in emission.
We restrict our investigation to low mass stars, of spectral type M0
and later, and we also concentrate on identifying bona fide flares with high
probability (rather than accepting lower probability events which may not be flares).  Our previous photometric and spectroscopic studies 
of the M dwarfs in SDSS have revealed correlations of the active
fraction with spectral type (later-type stars are more likely
to be active) and with distance from the Galactic plane, such that stars
closer to the plane are more likely to be active, and statistically
are members of a younger population \citep{West2004, West2008}.  The combination
of these effects leads to the result that later-type stars maintain
their activity for a longer time, perhaps for nearly the entire
$\sim$10 Gyr lifespan of the Galactic disk, while early M dwarfs lose their
activity in less than 1 Gyr \citep{West2008}.  We wish to determine
whether the stars that flare follow the same trends as the active stars,
as would be expected from the standard model connecting flare emission
with magnetic reconnection.  The SDSS Stripe 82 data provide a unique
sample with which to investigate these effects, since previous flare
data have been obtained only during continuous monitoring of individual,
nearby stars.

The second goal is to provide an initial set of flare frequencies
as a function of spectral type and Galactic position to compare to 
our numerical simulations of flaring rates across the Galaxy 
\citep[these simulations are in progress and will be reported in][in prep]{Hilton09}.  Our previous
work on the mass and luminosity functions of the low
mass stars in the Galaxy \citep{Covey2006, Bochanski2007, 
Bochanski2007b, Covey2008} together with our knowledge of
the active fraction of M dwarfs as a function of Galactic height
and spectral type, allows us to predict how many stars should
be flaring at a given Galactic position during a given exposure time.
Many of the parameters of such a global flaring model can only
be roughly estimated at present, and the SDSS Stripe 82 sample will give us
the opportunity to test whether the model is achieving reasonable agreement
with the data.  

The structure of the paper is as follows.
We first describe the SDSS dataset, and our sample selection which is 
designed to choose only those
targets with trustworthy observations for analysis (\S 2).  We then
discuss in considerable detail our 
efforts to develop a flare variability index that can be applied
uniformly to the time-resolved photometry to identify candidate flaring 
epochs (\S 3).
The results of applying this flare variability index to the sample are
discussed in \S 4, where we investigate the flare rates, correlation
with spectral type, activity status, distance from the Galactic plane, and line-of-sight through the Galaxy.
We also compare our flare luminosities and frequencies with previous 
determinations from nearby stars, and compare the flare 
rates with simple model predictions.  In \S 5, we summarize our results.

\section{Data}\label{sec:data}
The Sloan Digital Sky Survey \citep{York2000, Stoughton2002, Pier2003, Ivezic2004} is a large
($\sim$10,000 sq deg), multicolor \citep[\emph{ugriz}; ][]{Fukugita1996, Gunn1998, Hogg2001, Smith2002, Tucker2006} photometric and spectroscopic survey centered on the
Northern Galactic cap.  The Sloan 2.5m telescope \citep{Gunn2006},
located at Apache Point Observatory, operates in drift-scan mode as the 
camera \citep{Gunn1998} simultaneously images the sky in five bands, yielding photometry of 
$\sim$287 million unique objects contained in 
the SDSS Data Release 6 \citep{AM2008}. 

SDSS Stripe 82 is a narrow band comprising $\sim$300 square degrees along 
the Celestial 
Equator $(-51 \degr < \alpha < +60 \degr$, $-1.266 \degr < \delta 
< +1.266\degr$) that was observed more than twenty times 
under photometric conditions between 2001-2005.  Many additional ($40-60$)
SDSS-II imaging observations of Stripe 82 were obtained during
2005-2007; these data were not necessarily photometric and
required additional processing \citep{Ivezic2007}. The catalog of 
photometric observations we use incorporates the
\cite{Ivezic2007} calibrations, and was assembled as a queryable time-domain database at the University
of Washington in Spring 2008, in advance
of the date when the repeat 
Stripe 82 scans were released to the public as part of DR7 (Oct 2008, Abazajian et al. 2008).

\subsection{Sample Selection}\label{sec:sample}

The following steps outline the process we used to identify a sample of
M dwarfs from the SDSS Stripe 82 single epoch photometry (DR6), 
obtain their light curves (from the UW time-domain database,
though these data are now available in DR7), and restrict
the analysis to include only reliable, well-determined photometry.
Note that in this section and in \S 3, we perform all analysis
on the actual measured magnitudes (or fluxes) and their associated
errors.  In other parts of
the paper when considering color-spectral type relations, distances,
and luminosities, we apply interstellar reddening and extinction corrections to the measured
magnitudes using the standard SDSS corrections contained in the photometric
database, which are based on the dust maps from \cite{Schlegeldust}.

\begin{figure}
    \centering
    \plottwo{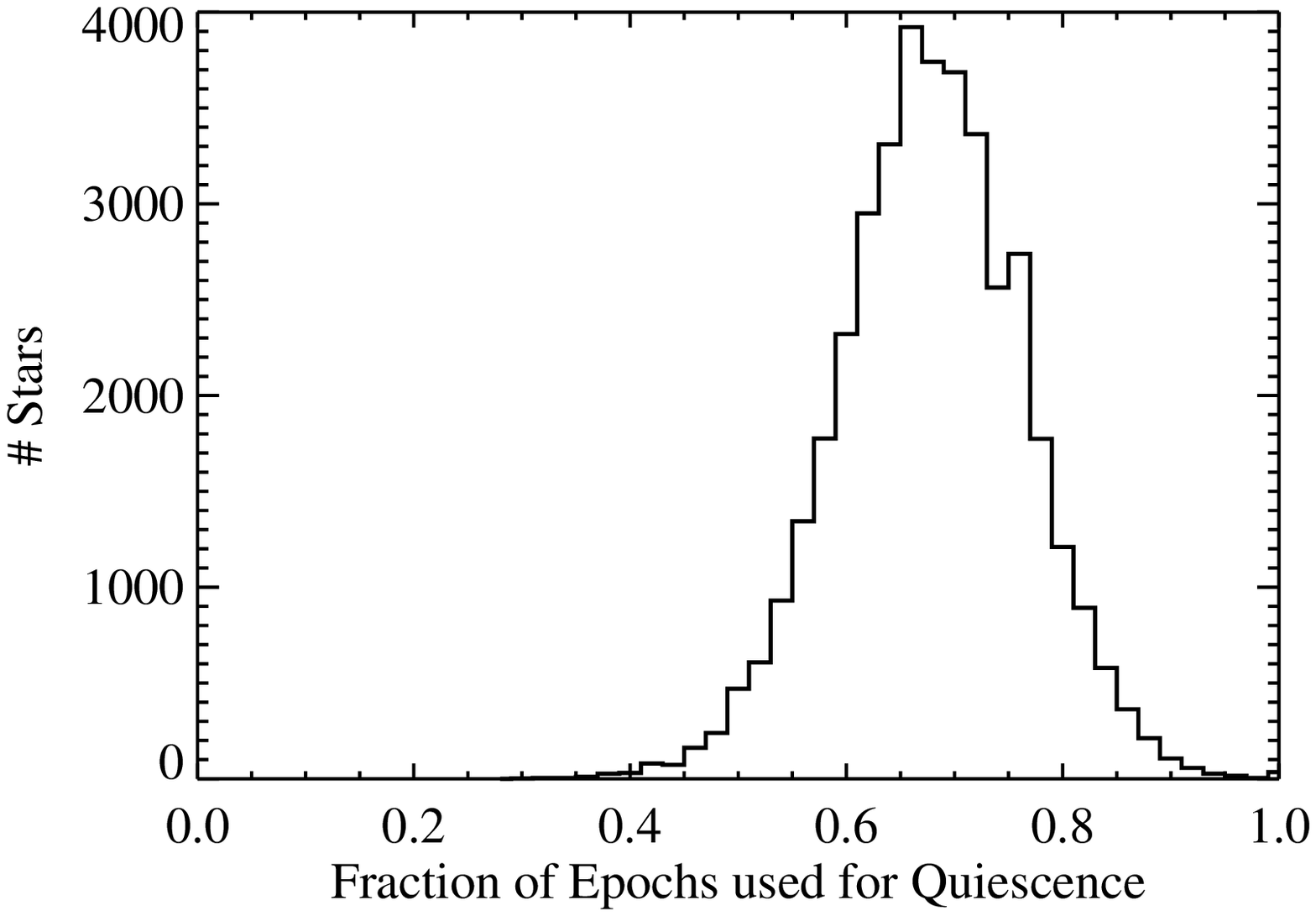}{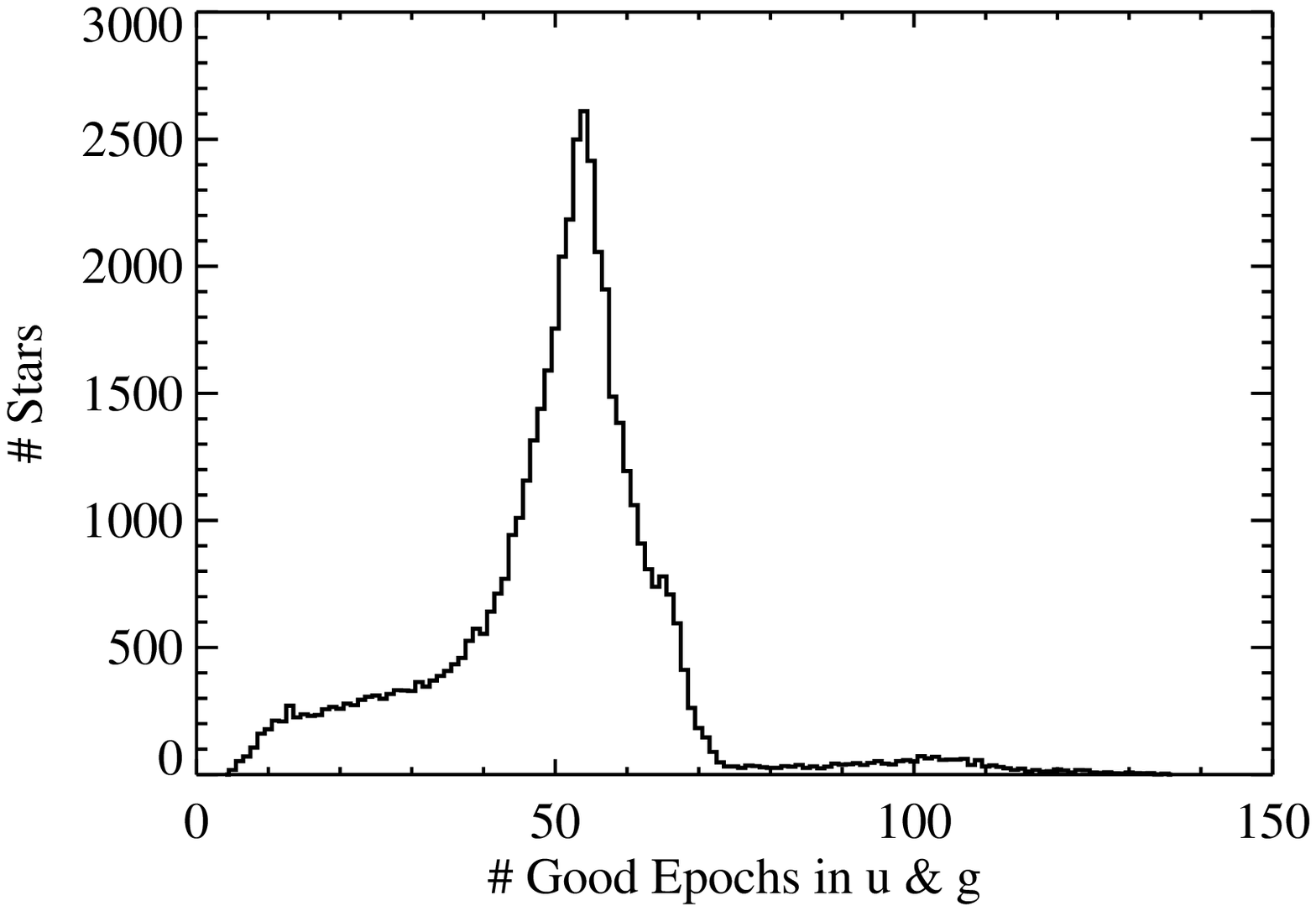}
    \caption{  Left -  The fraction of 
epochs used for the quiescent magnitude calculation shows that a 
significant number of observations were excluded 
due to being $> 3$ median absolute deviations away from the median.  
Right - The number of good epochs per star in both the  $u$ and $g$ filters.}
\label{fig:numeps}
\end{figure}

\begin{enumerate}

\item The SDSS Data Release 6 was
searched for point sources with the colors of low--mass stars in Stripe 82.  Based on 
\cite{West2005} and \cite{Bochanski2007}, the color limits were
chosen as $(r-i) > 0.53$ and $(i-z) > 0.3$ to restrict the sample to M stars.
The initial sample comprised more than 1.4 million candidate objects.

\item The photometry in each of the $r,i,z$ filters used to form the
colors was required to be brighter than the
survey limiting magnitudes.  This reduced the sample to
$\sim$900,000 candidate objects.

\item Standard SDSS photometry processing flags \citep{EDR} were used to reject objects due 
to poor DR6 single-measurement photometry.  In particular, data were removed if any of the
following flags were set:
SATURATED, NODEBLEND, NOPROFILE, PSF\_FLUX\_INTERP, 
BAD\_COUNTS
 \_ERROR,
INTERP\_CENTER, DEBLEND\_NOPEAK, NOTCHECKED.  The sample now
contained $\sim$700,000 objects from one of the following
SDSS Legacy runs: 2583, 2659, 2662, 2738, 3325, or 3388.

\item By far the most significant cut was the next one, where
the $u$ photometry was required to satisfy a brightness limit
$u < 22.0$.  The reason is that flares are most prominent in the
$u$-band, and if we cannot measure the quiet (outside of flaring)
$u$ magnitude accurately then we cannot compare the properties
of flaring epochs relative to the quiet star.  Because M dwarfs
typically have $(u-r) > 3$, this effectively restricts the
sample to stars with $r < 19$, which significantly decreases
the volume over which the stars can be observed.  Only $\sim$42,000
stars passed this cut.

\item The UW time-domain database was then employed to search for 
matches within $2 \arcsec$ of the DR6 position.  Observations were
kept if they had good astrometry, magnitude errors $<$ 1 mag, and good 
field-wide zeropoint measurements (important for data obtained under
non-photometric conditions).  Only a few hundred stars failed this cut.

\item To determine the mean quiet (non-flaring) 
magnitude in each filter for each star, 
the measured magnitudes were first
converted to fluxes (see \S 3.2 below), and a median flux was 
calculated for each filter.
A $3 \sigma$ (where $\sigma$
is the median absolute deviation from the median) cut was then made, and
the weighted mean flux (where each measurement is weighted inversely by the
square of the photometric error) was determined for each filter, subject to the requirement that at least five epochs 
had photometry within $3 \sigma$ available in each filter.
Mean fluxes were formed for $\sim$41,000 stars. 
Figure \ref{fig:numeps} (left)
plots the fraction (typically $>$ 60\%) of the total number of photometric 
observations that was used to determine the mean flux for each star.
Figure \ref{fig:numeps} (right) shows the number of good observations
in the $u$ and $g$ filters (that is, the observations returned from the 
UW time-domain database as per step 5 above, but not necessarily good 
according to the SDSS photometry flags).  The leftmost asymmetry is 
caused by stars that 
have $u > 21.7$ and therefore straddle the detection limit in the $u$-band.  
The peak near 100 observations is due to the stars that lie in the overlap
regions of camera columns between the N and S components of Stripe 82, and
therefore were observed twice as many times.

\item To ensure that the colors calculated from the light curves
are consistent with the initial DR6 color cut (see step 1 above), 
the same color cut was applied to the mean quiet
magnitudes, reducing the sample by a few hundred stars.\footnote{After correcting
for interstellar reddening and extinction the limiting $r - i$ color was reduced to 0.43 and the limiting $i - z$ color to 0.23.}

\item
A color cut requiring that $u-g > 1.8$ \citep{Smolcic2004} was
applied to eliminate $\sim$600 potential unresolved
white dwarf-M dwarf binaries, which have bluer $u-g$ colors.  The frequency of M dwarf - M dwarf binaries is likely $\sim$35\% \citep{FischerMarcy1992, Reid2002}.  Unresolved M dwarf binaries add uncertainty to the analysis, since the color of the star may not give an accurate quiescent luminosity and distance.  However, it is not possible to eliminate unresolved M dwarf - M dwarf binaries using photometry alone. 
\emph{Our final ugriz sample consists of 39,633 stars with 
observations totalling 1,987,253 epochs.}

\item
An additional photometric sample was obtained by following the prescription above
for stars with reliable $ug$ photometry, but $riz$ photometry that 
is saturated, or faulty in another 
way.  Starting with $u < 22, u > g$, and more stringent magnitude requirements
for the $u$ and $g$ bands (magnitude errors $\le 0.1$), an additional $\sim$500,000 point 
sources were obtained from the DR6 catalog in Stripe 82.  
Matching to 2MASS within $2 \arcsec$, requiring
good $K$-band photometry (SNR $>$ 5, magnitude error $<$ 0.1)
and applying the color cuts $g - K > 4$ and $u - g > 1.8$ leaves $\sim$12,000 stars.\footnote{After correcting
for interstellar reddening and extinction the limiting $g - K$ color was reduced to 3.6.}  
Matching to the time-domain database within $2 \arcsec$, and requiring
well-defined quiet magnitudes (see step 6, above), gives
\emph{the ug-only sample of 10,497 stars (524,981 epochs).} 

Although we initially analyzed the two samples separately, the results were
very similar and showed no systematic bias due to the lack of $riz$ colors
for the \emph{ug}-only sample.  Therefore, the rest of the analysis is shown
for the combined sample.

\end{enumerate}

\subsection{Spectroscopic Observations and Spectral Types}\label{sec:specobs}
About 10\% of the stars (4005, comprising 197,867 epochs) in the \emph{ugriz} photometric 
sample have SDSS spectroscopic observations, determined by matching to the spectroscopic 
DR6 database. None of the \emph{ug}-only sample have SDSS spectra (these
stars had poor \emph{riz} photometry and therefore were likely rejected by the spectroscopic
targetting algorithm).  The stars with SDSS spectra have been assigned
types using the Hammer spectral typing facility \citep{Covey2007} and have been
previously analyzed for magnetic activity 
based on detecting emission in the chromospheric H$\alpha$ line
\citep{West2008}.
Stars characterized as \emph{active} and \emph{weakly active}
by \cite{West2008} comprise the active sample (219 stars), with
the rest of the spectroscopically observed stars being labeled inactive (3786 stars).

The vast majority of stars in our sample do not have spectra, so
we assign an estimated spectral type based on the photometric
colors.  The estimated spectral types serve as a proxy for mass and surface
temperature and provide a convenient way to subdivide the sample
into intuitive ``early, mid, late'' spectral type bins to guide the analysis and 
compare to M dwarfs which do not have SDSS photometry.
The colors used to assign types must first be corrected for extinction
in order to avoid the common problem of assigning a reddened, early-type
star at large distance an erroneous later spectral type, which would in turn
place it at an (incorrect) closer distance.
Since Stripe 82 is located in the Southern Galactic hemisphere, our line
of sight to nearly all stars (except the very closest) passes through
the plane of the Galaxy. We therefore account for the interstellar reddening and
extinction as described in \S2.1.

\begin{figure}
    \centering
    \includegraphics[height=0.4\textheight]{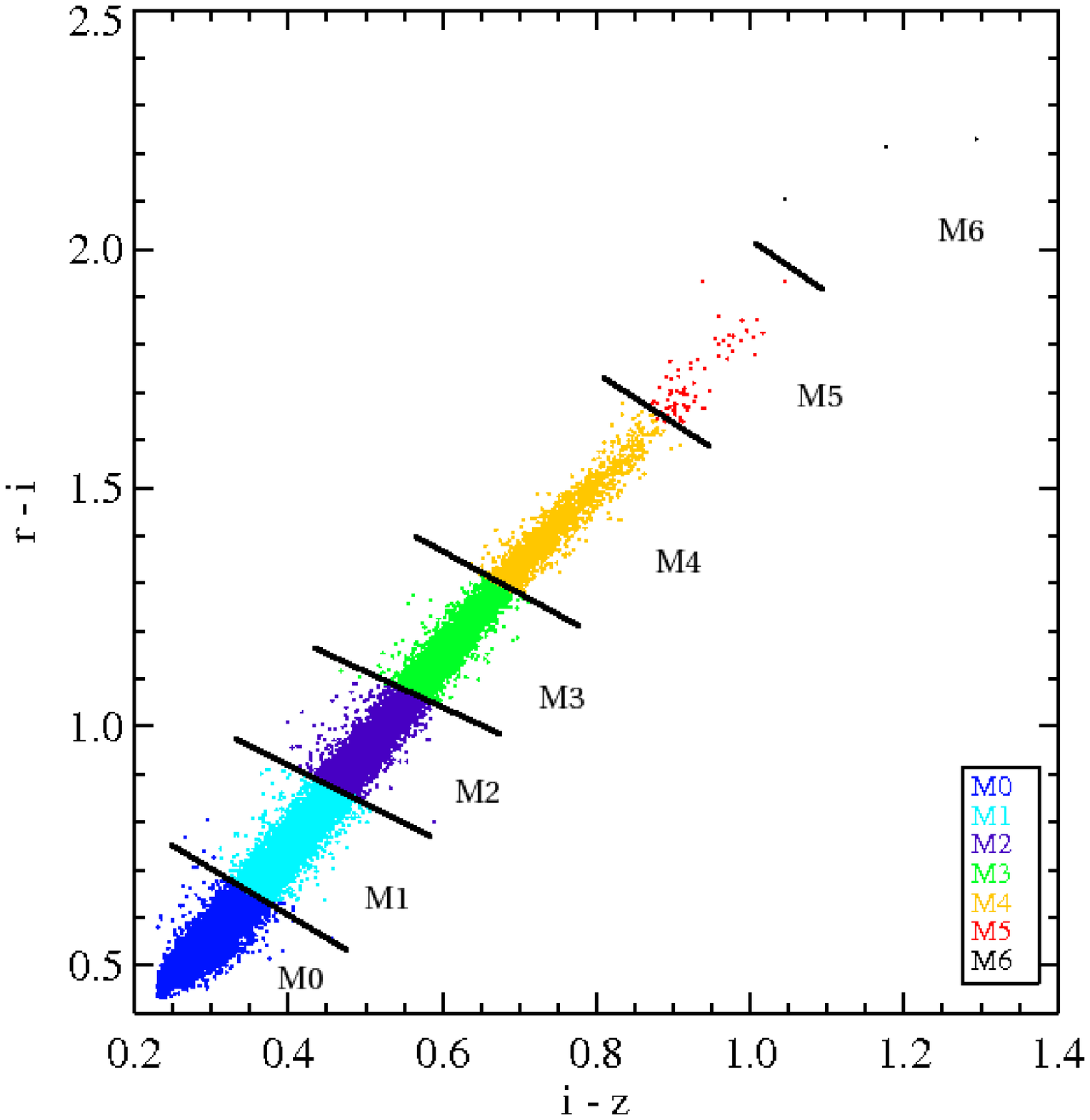}
    \includegraphics[height=0.4\textheight]{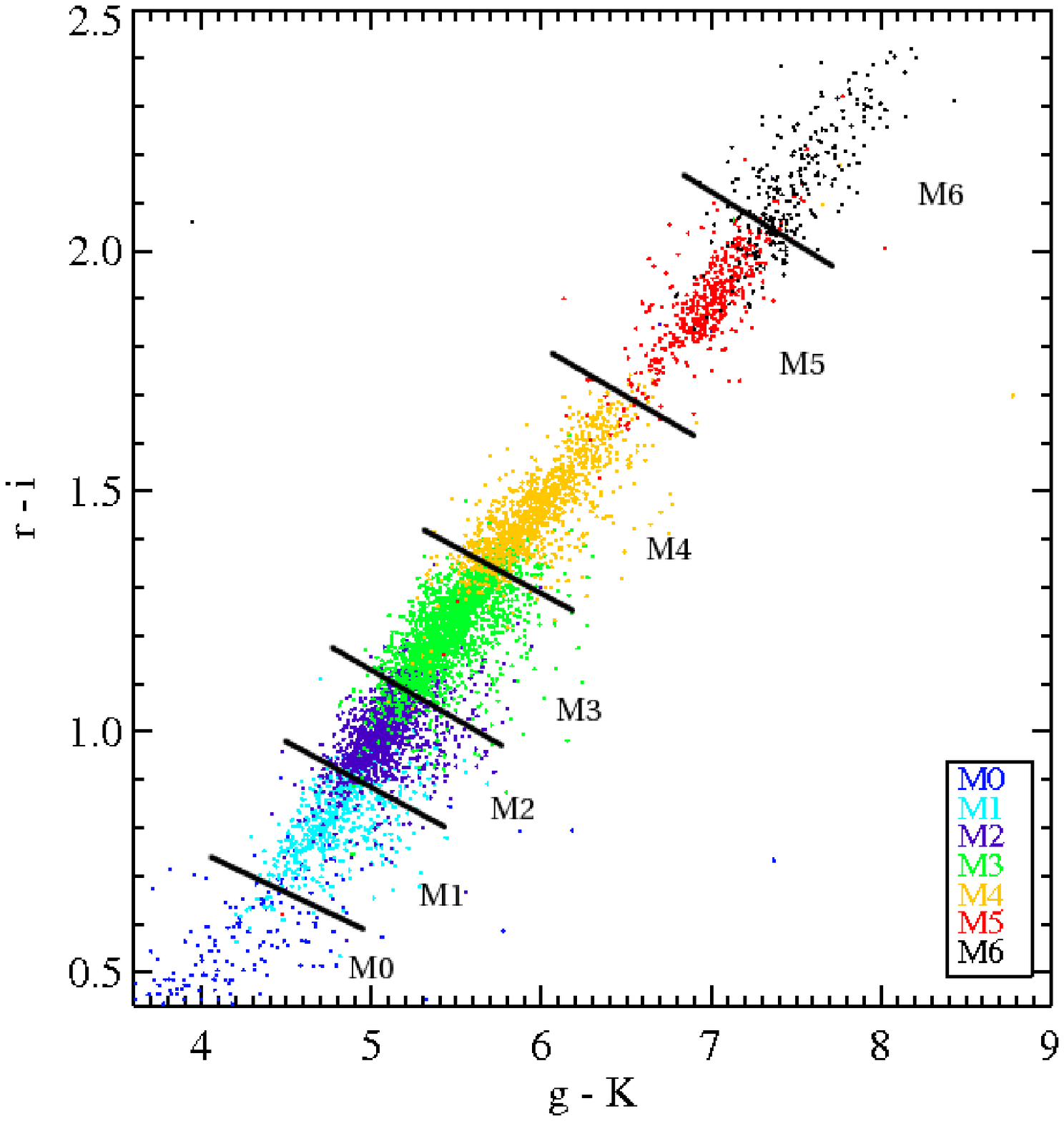}
    \caption{Top - The extinction-corrected $r - i$ vs. $i - z$ color-color diagram for the sample of $39,633$ M dwarfs with well-defined light curves.  
The spectral types were probabilistically assigned based on our presented spectral type-color relations (Table \ref{table:photclass_param}).  Bottom - 
The extinction-corrected $r - i$ vs. $g - K$ distribution of $\sim$7,200 
stars with good photometry from the SDSS DR5 spectroscopic sample \citep{West2008}.
Using this sample, we derive the mean and standard deviation of $g - K$ 
for each spectral type, which allows us to
estimate the spectral types of the $ug$-only sample.}
\label{fig:colors}
\end{figure}

The ($r - i$, $i - z$) color-color diagram was used to classify the \emph{ugriz} sample.  
We fit 2D Gaussian distributions using FastMix \citep{Moore99} to the 
\cite{West2008} DR5 M dwarf spectroscopic sample comprising some 20,000 stars with good $riz$ colors
and accurate spectral types based on SDSS spectra.\footnote{We augmented the original DR5 sample with
additional M0 stars having bluer colors in order to fit the full color range
of the extinction-corrected colors in our sample}
The resulting best fit parameters (mean and covariance matrices, given in
Table \ref{table:photclass_param}) define a 
2D Gaussian probability function for each spectral type, allowing us
to assign estimated spectral types based on the photometric colors, as shown in the top panel of Figure \ref{fig:colors}.

For the $ug$-only sample, we estimate the spectral types according to 
the $g - K$ color \citep{Covey2008}.  Again using the DR5 M dwarf spectroscopic 
sample, with $K$ magnitudes
from 2MASS, a 1D Gaussian probability function for the 
extinction-corrected $g - K$ color was determined 
for each M spectral type.  The bottom panel of 
Figure \ref{fig:colors} shows the
distribution of $r - i$ vs. $g - K$ 
indicating that both colors provide reasonable estimates of the spectral type.

\begin{figure}
    \centering
    \includegraphics[height=0.5\textheight]{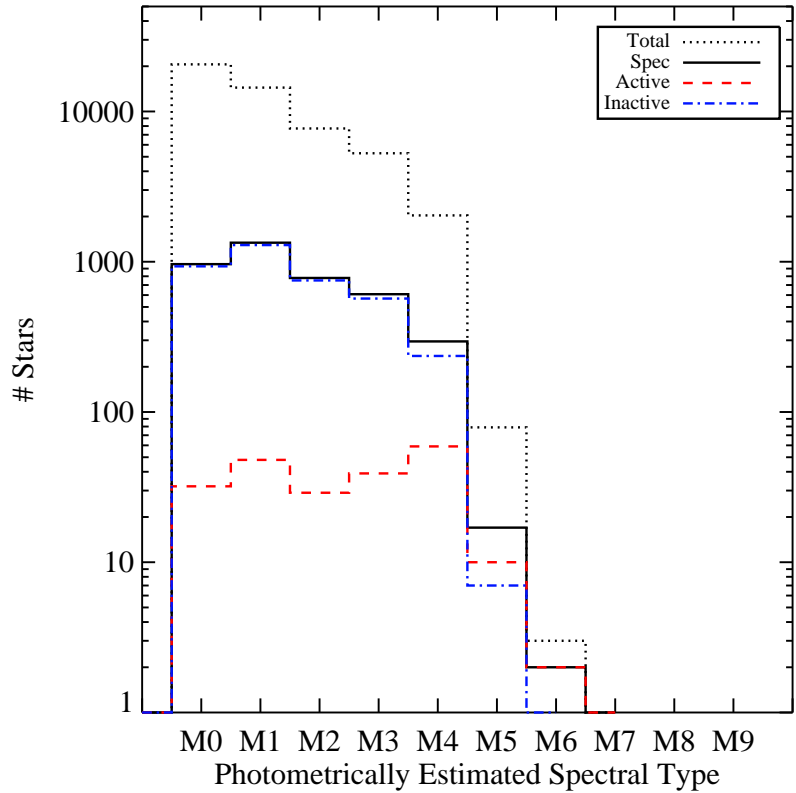}
    \caption{ The distribution of M dwarfs sorted by the photometrically estimated spectral type for 
the total photometric sample ($50,130$ stars), the 
total spectroscopic sample ($4,005$ stars), the active stars ($219$ stars), and the 
inactive stars ($3,786$ stars).  }
\label{fig:hist_all}
\end{figure}

The dotted line in the top panel of 
Figure \ref{fig:hist_all} 
gives the estimated spectral type distribution of the photometric 
(combined \emph{ugriz} and \emph{ug}-only)
sample.  The effect of the limiting apparent magnitude of the sample 
($u < 22$) is clearly evident through the relatively large numbers of 
early-type (M0-M1) stars which can be seen to 
larger distances, and the small numbers of later-type (M4+) stars which 
can only be observed
nearby.  Therefore, the later-type stars come from a much smaller volume
than the early-type stars.  The colored lines in Figure \ref{fig:hist_all} 
show the distributions of active and inactive
stars as a function of spectral type.  As expected \citep{West2004},
most of the early-type M dwarfs are inactive, while most of the later
ones are active.

Finally, we investigated the contamination by giants in the sample.  
Separating giant and dwarf stars without high resolution spectra is 
problematic. \cite{Covey2008} examined the $J$ vs. $i - J$ 
color-magnitude diagram of 
two calibration regions (1 sq deg each) in Stripe 82.  We verified 
that the colors of the stars in our sample lie in the region of 
this color-magnitude diagram with $< 2$\%
contamination from giants.

\subsection{Followup Spectra from the ARC 3.5m Telescope}\label{sec:apo}

Time domain survey data potentially allows the investigation of flaring rates from
both active and inactive stars.  The former have been well studied for many years, but
the latter have received very little attention.  Because we now have available a
sample of ``stars that flared'', it is interesting to know if they are active or
inactive in their quiescent state.  Therefore we obtained followup spectra with 
the ARC 3.5m Telescope at Apache Point Observatory, using the Dual Imaging
Spectrograph (DIS) with the R300 grating giving wavelength coverage $\sim$6000 \AA$ - $9200 \AA.  On UT080825, we used a 
2.0$\arcsec$ slit giving R $\sim 800$ and on the dates UT080919, UT081013, UT081107, UT081124, 
and UT081214 we used a 1.5 $\arcsec$ slit giving R $\sim 1000$.  As 
these stars are quite bright in $r$, exposures of 3-15 min
gave S/N $> 10$ at H$\alpha$.  Blue spectra were also obtained; however, 
the S/N was inadequate to obtain information about the activity status 
of the other H Balmer lines or Ca II H \& K.   
The data were reduced using standard IRAF\footnote{IRAF is distributed by the National Optical 
Astronomy Observatories, which are operated by the Association of Universities for Research in 
Astronomy, Inc., under cooperative agreement with the National Science Foundation} procedures, wavelength calibrated using 
HeNeAr lamp spectra, and flux-calibrated using spectrophotometric standard 
star observations (though 
some of the data were taken during non-photometric conditions, limiting them 
to a relative flux calibration).
We discuss these observations further in the context of the relative
probability of flares on active and inactive stars in \S 4.6.

\section{Variability Analysis}\label{sec:FIanalysis}

\subsection{Properties of Flare Photometry}

The SDSS Stripe 82 photometry for each M dwarf typically consists of $\sim$50
good observations (Figure \ref{fig:numeps}), irregularly spaced in time with 
separation of several days, during the 
fall months, spanning several years. 
These data allow us to construct a 
low cadence light curve in each filter.
Since flaring timescales range from minutes to hours,
individual flares will be observed only once during their time-evolution.  Therefore, each epoch
of each star must be analyzed for a flare signature. 
An example of the \emph{ugriz} light curves for an active M6 dwarf is 
shown in Figure \ref{fig:lc}.  This star flared (with increases\footnote{Although increases in flux correspond to negative
changes in magnitudes, to avoid confusion we refer to flares as having positive magnitude changes $\Delta$$u = |u_{flare} - u_{quiet}|$ throughout the paper.} of 5.5 magnitudes in the 
$u$-band, and 3.1 magnitudes in the $g$-band) during the first epoch marked
with a star and again at much lower levels in the second marked epoch.  

Flare emission can change significantly over the timescale of the SDSS photometric
observations, which comprise $\sim$54 seconds in each filter as the star passes
across the detector, in the filter order \emph{r-i-u-z-g}.  For the $u$ and $g$ observations
which are primarily of interest here, this means that the $g$ observation begins
108 seconds after the $u$ observation, and there is a 54 sec gap between them.
Figure \ref{fig:adleolc} illustrates the photometry measurements for a small flare, with a 0.6 mag increase at the peak in the $U$-band (similar
to $u$) and a 0.1 mag increase in the $B$-band (similar to $g$), obtained with the NMSU 1m telescope during April, 2008 \citep[in prep]{KoKo09}.  Depending
on the timing of the SDSS observations obtained during such a flare, our requirement that
both the $u$- and $g$-bands show enhancements (as described in the next section) 
would inefficiently select flares of this size.
This is one of the reasons that we limit our flare detection capability to only those
events that show 0.7 mag or larger increases in the $u$-band (see \S 3.3). 

\subsection{Flare Variability Index}  
We have developed a Flare Variability Index, 
$\Phi_{ug}$, that
is a modified version of an index originally designed 
to measure variability in multi-epoch and multi-band photometry \citep{Welch1993, Stetson1996}. The Welch and Stetson Variability Index is often 
employed for periodic variable searches, e.g., for Cepheids and RR Lyrae stars.
Our modification produces a variability index that is calculated 
on an epoch-by-epoch basis.
The Flare Variability Index, $\Phi_{ug}$, is based on measuring positive flux increases in both the
$u$ and $g$ filters, 
since the defining characteristic of stellar flares is the
dramatic enhancement in the blue continuum emission \citep{Moffett1974}.
Note that the photometric data in the UW time-resolved database are given in
magnitudes, which we convert to nanomaggies (nMgy), a linear
flux density unit, via Equation 1 in \cite{Ivezic2007}.
One Mgy corresponds to a flux density of $3631$ Jy \citep{Oke1983, Finkbeiner2004, 
Blanton2005}.  With fluxes expressed in nanomaggies,
we quantify the increases in the \emph{u} 
and the \emph{g} fluxes at a given epoch on a given star by comparing them 
with the measurement errors:

\begin{equation}
 \Phi_{ug} = \frac{(F_{u,j}-F_{u,quiet})}{\sigma_{F_{u,j}}} \times \frac{(F_{g,j}-F_{g,quiet})}{\sigma_{F_{g,j}}},
\end{equation}

\noindent where $\Phi_{ug}$ is the Flare Variability Index evaluated at 
epoch \emph{j},
$F_{u,j}$ is the flux in the $u$ filter at epoch \emph{j}, $F_{u,quiet}$ is
the mean $u$-flux for the star, and $\sigma_{F_{u,j}}$
is the photometric uncertainty in the $u$ flux at epoch \emph{j}; 
the same notation is used for the $g$ filter quantities. 
The quiet flux values $F_{u,quiet}$ and $F_{g, quiet}$ are taken
to be the weighted means of the well-measured epochs (see \S 2.1).

Flares will have positive flux deviations in both $u$ and $g$,
\begin{equation}
 F_{u,j}-F_{u,quiet} > 0; F_{g,j}-F_{g,quiet} > 0,
\end{equation}
generating a positive value of 
$\Phi_{ug}$.  Note that eclipses, which have
negative flux deviations in both filters, also generate
a positive value of $\Phi_{ug}$, while the subset of random noise characterized by positive
flux deviation in one filter, negative in the other, generates
a negative value of $\Phi_{ug}$.  To insure that eclipses
are not included as candidate flare events, we require that
the $u$-band flux enhancement be positive.

\subsection{Flare Candidate Selection}
\begin{figure}
    \centering
\includegraphics[height=0.9\textheight]{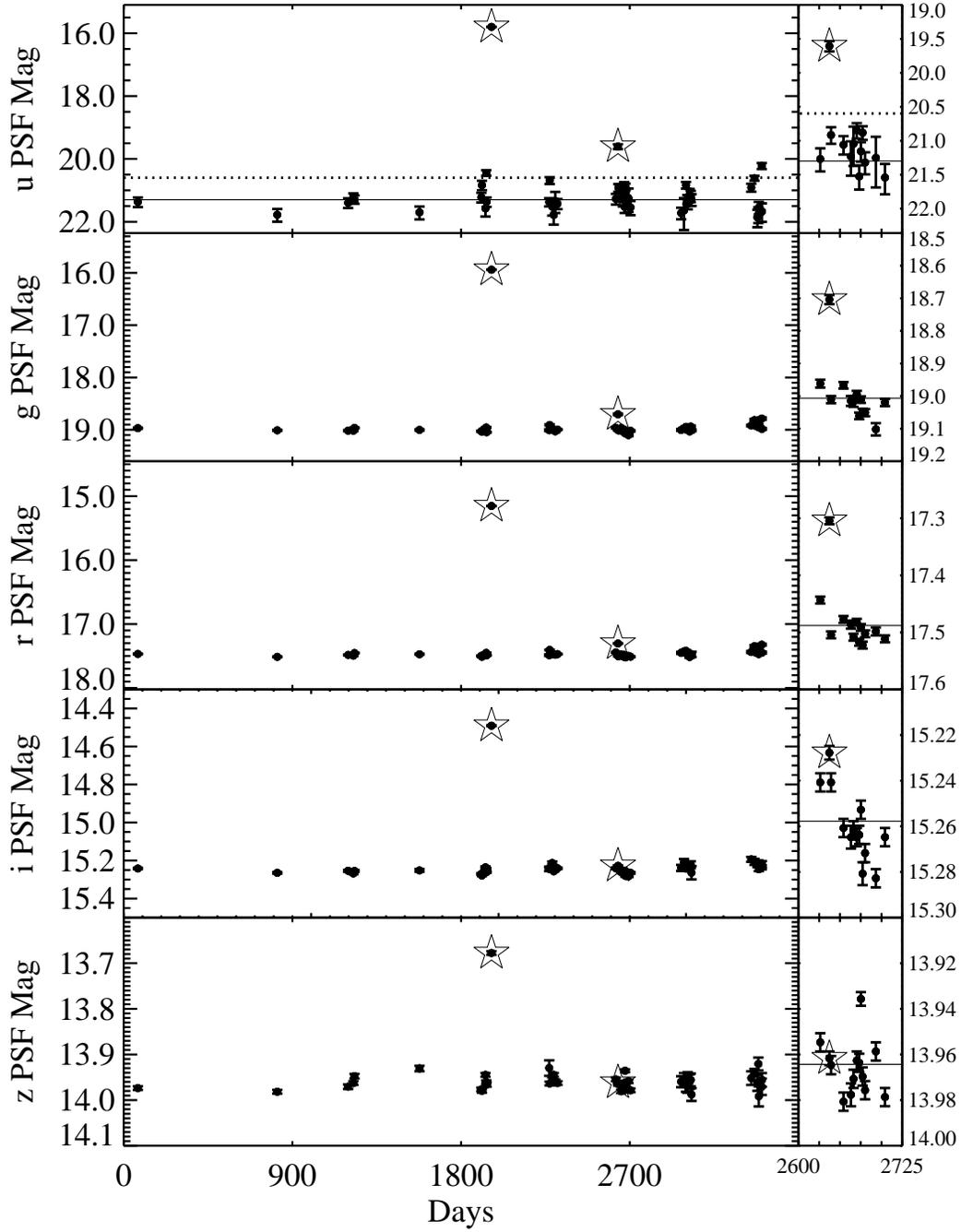}
    \caption{An example light curve with $46$ observations in $ugriz$ 
for an active 
M6 dwarf which exhibits both a large and small flare in the Stripe 82 dataset (highlighted with star symbols).  The magnitude enhancements for the large flare are $\Delta$mag$ = 5.50, 3.07, 2.34, 0.77, 0.29$ in $ugriz$, respectively.  Note that the small flare shows negligible enhancements in $i$ and $z$. The inherent large 
variations in the $u$-band necessitate our requirement for a simultaneous
enhancement in the $g$-band. The panels to the right zoom in on 13 observations (over 78 days) 
around the small flare.  The solid line indicates the quiescent magnitude whereas the dotted line
shows the 0.7 magnitude flare detection threshold.}
\label{fig:lc}
\end{figure}

The quiet flux of M dwarfs in the \emph{u}-band is very faint, in most
cases close to our apparent magnitude limit of $u = 22$.
Thus we expect that 
photometric variations caused by random sampling of noisy data 
(e.g. the significant noise apparent in the $u$ and $g$ light 
curves in
Figure \ref{fig:lc})
will in some cases produce positive values of the Flare 
Variability Index, and hence masquerade as flares.  
The False Discovery Rate (FDR)
method of \cite{Miller2001} allows us to further limit the number of positive
$\Phi_{ug}$ events that are candidate flares.  The FDR analysis involves
comparing a source distribution with a null distribution, and then using the 
null distribution to assign
a probability that a given value of $\Phi_{ug}$ is drawn from the null.
Figure \ref{fig:FIdist} compares the candidate flare (source) distribution 
($\Phi_{ug} > 0$, $F_{u,j}-F_{u,quiet} > 0$, $F_{g,j}-F_{g,quiet} > 0$), with the random noise (null) distribution
($\Phi_{ug} < 0$), for the same number of epochs in each distribution ($\sim$770,000).
It is clear that a significant number of the candidate flares, especially
at low values of $\Phi_{ug}$, can be attributed to random noise.  False Discovery
Rate analysis
provides a way to select a threshold $\Phi_{ug}$ value that limits the percentage of 
false positives.
Employing the IDL routine in Appendix B of \cite{Miller2001},
we determined that setting the FDR parameter, $\alpha$, to $10\%$ corresponds to $\Phi_{ug} \sim 100$
(the vertical line in Figure \ref{fig:FIdist}).  Therefore, no more than 10\% of the epochs with $\Phi_{ug} \ge 100$ are false positives caused by random sampling of noisy data.
This returned a reasonable number (673) of candidate 
flare epochs for individual examination. 

\begin{figure}
    \centering
    \includegraphics[height=0.4\textheight]{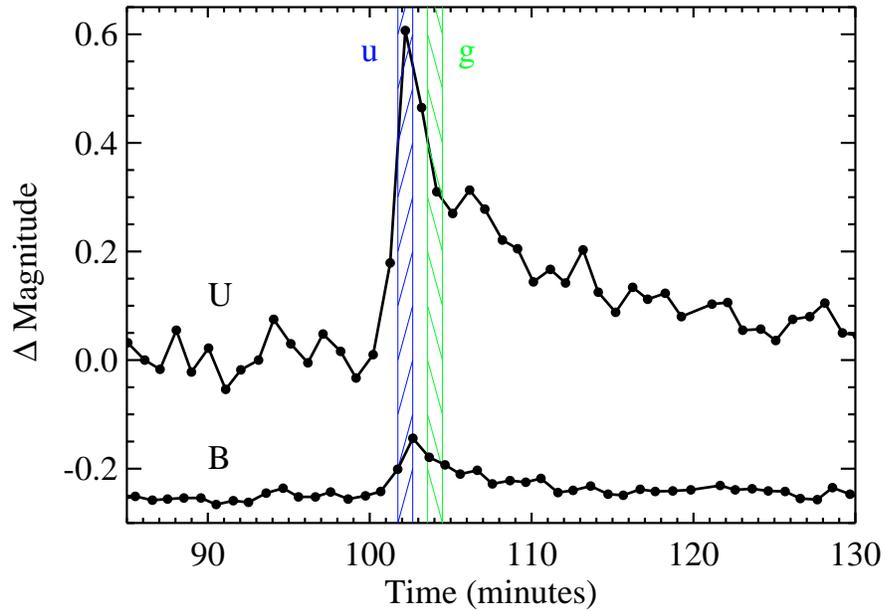}
    \caption{Example light curves in $U$ and $B$ of time-resolved broadband photometry during a 
$\sim$ 0.6 $U$-band peak magnitude flare \citep[in prep]{KoKo09}.  The time-sequence 
of the SDSS $u$ (blue) and $g$ (green) filters are shown as the hashed regions
in the unlikely scenario that the $u$-band integration caught the flare 
exactly at the peak.  The time 
sequence introduces uncertainty in the interpretation of the SDSS 
flare emission, especially for small, short-lived flares. }
\label{fig:adleolc}
\end{figure}

\begin{figure}
\centering
    \includegraphics[width=0.75\textwidth]{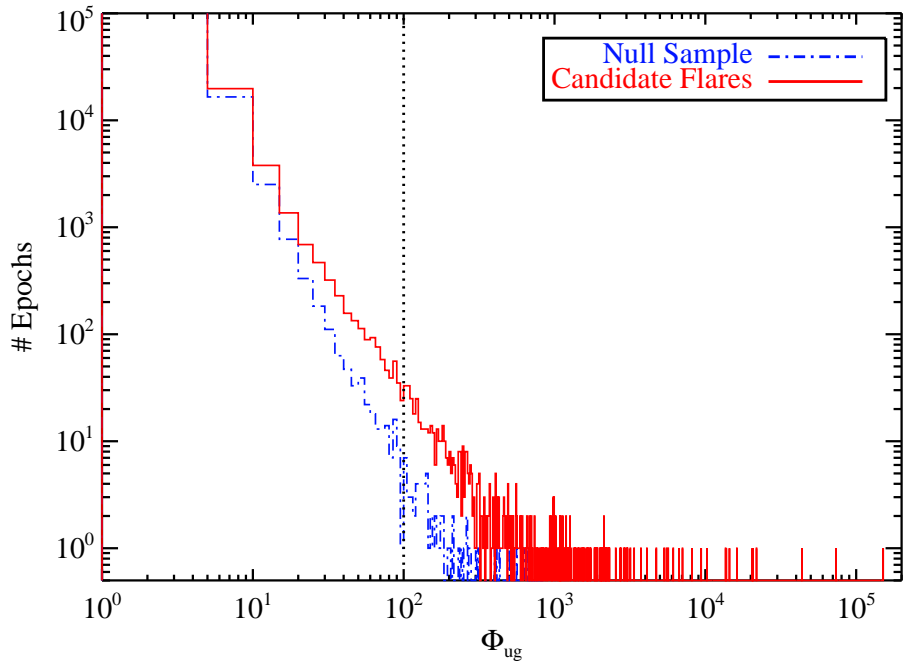}
 
    \caption{The distribution of the flare variability index for our candidate flare distribution
(red curve; $\Phi_{ug} > 0$, $F_{u,j}-F_{u,q} > 0$, and
$F_{g,j}-F_{g,q} > 0$) compared to the 
distribution of $\Phi_{ug} < 0$ (blue curve) shown for the same number 
of epochs ($\sim$770,000).  The former comprises both candidate flares and random
night-to-night photometric variation, while the latter results purely from random photometric variation. The vertical line indicates the threshold value of $\Phi_{ug} = 100$ which provides $< 10$\% false discoveries, and gives
$673$ candidate flares (see text).  }
\label{fig:FIdist}
\end{figure}

\begin{figure}
    \centering
    \includegraphics[width=0.5\textwidth]{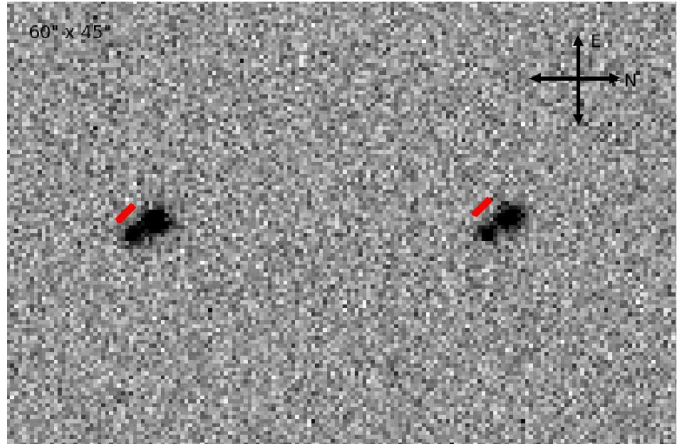}
    \caption{ An example of the red leak in SDSS $u$-band images for two 
red stars (both $g - r = 1.44$) during a non-flaring epoch 
(run $ = 6590$, camcol $ = 2$, field $ = 218$) at a high airmass 
of 1.64.  These represent the worst cases of red leak encountered, with
nearly complete separation of the red lead and true stellar images.
A line at the parallactic angle with length equal to the displacement 
of $\sim$2.3\arcsec \ predicted from \cite{Filippenko1982}
is overplotted in red next to each star.
If all the red leak flux were added to the $u$ flux 
and not deblended, 
the red leak would contribute approximately $0.5$ magnitudes to the
photometry for each star. }
\label{fig:redleak_image}
\end{figure}

As shown in Figure \ref{fig:adleolc} and described briefly in \S 3.1, 
flares with small flux enhancements typically occur over short timescales and
may not be detected in both the $u$ and $g$ filters during the SDSS
imaging cadence.  An additional complication is that the SDSS $u$-band imaging suffers from a photometric defect known as
the ``red leak'' caused by the $u$ filter which allows flux with wavelength longer than 7100\AA\ 
to be partially transmitted.
Due to the complexity and time dependence of this instrumental artifact,
it is not corrected for in the SDSS photometry.\footnote{see http://www.sdss.org/dr7/products/catalogs/index.html for a summary description of the red leak}
We investigated the red leak by analyzing SDSS images taken at
a range of airmasses. The differential refraction between the blue and red
light separates the red leak photons into a faint image that
is offset from the (blue, $u$-band) stellar image at the parallactic angle
and with the correct image displacement, ranging from 1.15\arcsec to 2.4\arcsec \citep{Filippenko1982}.
As a result, there may be various degrees of PSF deblending from observation to
observation depending on the seeing, airmass, and stellar and flare colors.  
We show examples
of the red leak effect in Figure \ref{fig:redleak_image}.  In the two worst 
cases, we estimate that the amount of the red leak would have added 
$\sim$0.3 and 0.5 mags to the $u$-band measurements had they been 
co-aligned and therefore not deblended.  Thus, we do not
trust any measured $u$-band magnitude enhancements of 0.5 magnitudes or less.
Since, as described in the Introduction, we only want to analyze bona fide flares, we adopt a 
conservative limiting magnitude of
$\Delta u = |u_{flare} - u_{quiet}| \ge$ 0.7 magnitudes, which corresponds 
to almost a factor of two increase in the flux,
as the lower limit of our flare detection capability.  Requiring a 
correlated increase in \emph{g} is an additional safeguard against erroneously selecting ``flares'' due to 
the red leak.

Flares typically show blue
continuum emission with the approximate spectral shape of a 10,000K
blackbody (\cite{HF1992}, see also Figure \ref{fig:adleolc}).   This characteristic
implies that the flux enhancement in $u$ should be
greater than that in $g$, 
\begin{equation}
  e_u = \frac{F_{u,j}}{F_{u,quiet}} > \frac{F_{g,j}}{F_{g,quiet}} = e_g
\end{equation}
All but two of the candidate flares exhibit this characteristic.  
The time evolution 
of the flares combined with the image sequencing adds ambiguity to the 
precise spectral shape of the flares, and so we do not exclude 
candidate flares that fail to exhibit this characteristic.  In
rare cases, observations during the rapid rise phase of the
flare, if the $u$-band was measured just as the flare started, and 
the $g$-band slightly later, at the peak of the flare, might show
$e_g > e_u$.

The number of candidate epochs is now small enough that each can be queried
in the time-domain database to return photometry flags, as was originally
done for the SDSS DR6 Stripe 82 photometry (see \S 2.1).
The following flags cuts were applied in the $u$ and $g$ bands to eliminate observations
with poorly determined photometry:
SATURATED, NODEBLEND, NOPROFILE, PSF\_ FLUX\_INTERP, BAD\_COUNTS\_ERROR,
INTERP\_CENTER, DEBLEND\_NOPEAK, NOTCHECKED, MAYBE\_CR and SATURATED $= 0$.  
Note that most of the rejected epochs resulted from having NODEBLEND set,
indicating that a star is seen as two fainter objects in good seeing,
but a brighter object (masquerading as a flare) in poor seeing, when the two objects cannot
be deblended.

The candidate flare sample has now been pared as much as possible by
statistical analysis, noise requirements, and flare physics.  Further
analysis requires visual inspection of the data for each epoch.
An image was retrieved and examined to look for artifacts such as 
diffraction spikes.
The light curves were also examined to insure that the data 
are consistent with those seen from flares
(e.g. Figure \ref{fig:lc}).
Finally, to confirm the increases in flux indicated by the automatic
SDSS PSF photometry,
we identified a non-flaring epoch for each star that had a candidate flare
epoch, and ran a PSF-matching and image subtraction algorithm developed for the SDSS-II
Supernova Survey and the Deep Lens Transient Survey \citep{Becker2004}.  
During a true flare, 
the field stars will disappear in the subtracted images, and the 
flare star will show residual PSF structure in both the $u$ and $g$ images.  
For cases with unusual PSF structure or data taken with poor seeing, we calculated
the aperture photometry for the flare star and compared to several other stars in the field, as
the aperture photometry should be independent of the PSF shape. Only 9 flare candidates
were eliminated through this visual inspection, image subtraction, and aperture
photometry process.

\begin{figure}
\centering
    \includegraphics[width=0.75\textwidth]{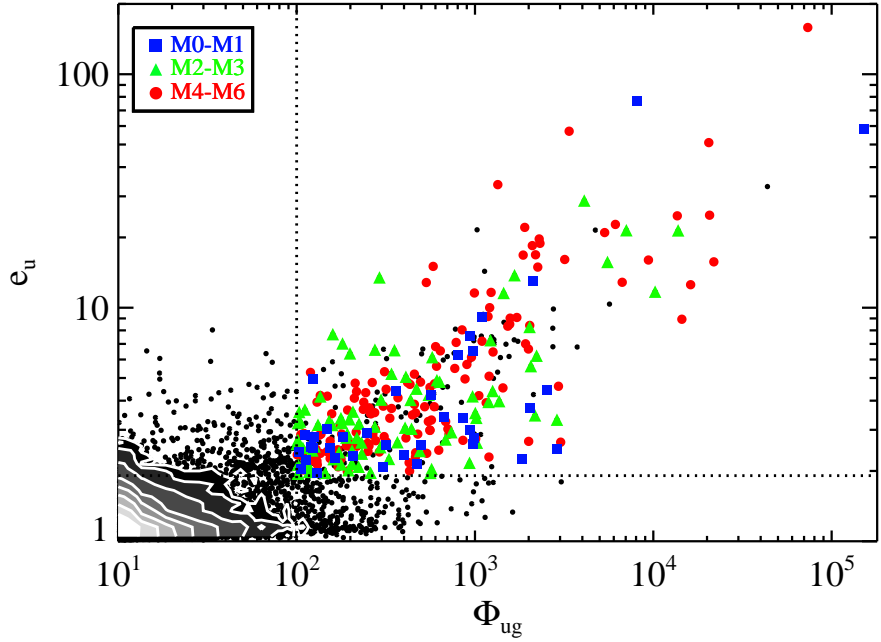}
  \caption{Top - The flux enhancement in the $u$ filter vs. $\Phi_{ug}$ for $\Phi_{ug} > 0$,  $F_{u,j}-F_{u,q}> 0$, and $F_{g,j}-F_{g,q} > 0$.  
The vertical line corresponds to critical threshold $\Phi_{ug} = 100$ and the horizontal line
corresponds to a $\Delta u$ mag $= 0.7$.  The 271 flaring epochs are shown 
as colored circles, while the observations which are cut out by flags or 
image inspection are left as 
black circles.  The contours correspond to density levels of (5, 10, 25, 75, 150, 300, 600, 1200) in bins of
$\Delta \Phi_{ug} = $ 5, $\Delta e_u =$ 0.1.}
\label{fig:FI_dumag}
\end{figure}

Figure \ref{fig:FI_dumag} shows the selected flare candidate region in
a plot of $u$-band flux enhancement versus Flare Variability Index.
Black points show 
all observations with $\Phi_{ug} > 0$,  
 $F_{u,j}-F_{u,q} > 0$, 
$F_{g,j}-F_{g,q} > 0$.  
The vertical line corresponds to $\Phi_{ug} = 100$ and the horizontal line
corresponds to $e_u = 1.91$ ($\Delta u$ mag = 0.7).  The 271 flare epochs are shown as circles, colored by most 
probable spectral type. To the left of our  $\Phi_{ug}$ threshold, observations
may have a large flux enhancement; however they are not selected due to either a large
photometric error in $u$ or a negligible increase in $g$.  We inspected 
several candidate epochs that fell just short of making our cuts, and found
that the return in actual flare observations was very small compared to 
the increase in the number of candidate epochs resulting from weakening
our criteria.

\subsection {Summary of Flare Sample}

Table \ref{table:steps_ugriz} summarizes the steps we took to arrive 
at our final flare sample, which consists of 271 epochs (225 from the $ugriz$ sample
and 46 from the $ug$-only sample) on 236 stars. We verified that the fraction of observations that flare for all the runs is fairly uniform,
consistent with observing random stellar events and not the result of photometric
artifacts on one or a few nights.
 There are 38 flares (38 stars) on early M0-M1 dwarfs
out of 1,783,142 epochs (35,024 stars) observed; 
83 flares (79 stars) on mid M2-M3 dwarfs out of 629,170 epochs (12,986 stars) observed, and 
150 flares (119 stars) on later M4-M6 dwarfs out of 99,922 epochs (2,120 stars) observed.  Clearly
the flaring fraction is much higher on the later-type stars, both because they
are more likely to be magnetically active, and because flares at the level
of twice the quiescent brightness are much lower energy on fainter stars,
and therefore likely to be observed more often.  These results are discussed further
in \S \ref{sec:flare_fraction} below.  Note that several of the redder
stars flared multiple times, with
the highest occurrence being 5 flares (and possibly several more small flares that failed to make our cuts) on an active M5 star (SDSS J025951.71+004619.1). 

\section{Results \& Discussion}

\subsection{The Fraction of Stars that Flare}\label{sec:flare_fraction}

The Stripe 82 imaging data provided observations of 
$2,512,234$ epochs with reliable photometry in both the $u$ and $g$ bands, 
from which we confidently identified 271 flares.
This gives an overall flaring fraction of $0.0108\pm0.0007$\%. 
In other words, about 1 of every $10,000$ observations in Stripe 82 of a
star with the colors of an M dwarf and $u < 22$ resulted in a flare detection
with $\Delta u \ge 0.7$ mags.  Several factors enter into this determination
of the flaring fraction: 
the stellar density distribution being sampled to an apparent magnitude limit
of $u < 22$, resulting in a large number of early type M dwarfs in the sample; 
the fact that active stars are more likely to flare, and are more likely to
be of later spectral type and therefore are being sampled over a smaller
volume; and the flare detection limit of $\Delta u \ge 0.7$ magnitudes which
results in lower energy flares being detectable only on intrinsically fainter stars.
We discuss each of these effects separately.

As expected, the stars that were known to be active, based on SDSS spectral observations
showing H$\alpha$ in emission, flared much more frequently.
Out of 10,396 observations of known active stars, we
found 29 flares, which gives a flaring fraction of $0.28\pm0.05$\%, 
$\sim$30 times higher than the flaring fraction for the whole sample. 
There were no known (from SDSS spectra) inactive stars that flared (but see
\S 4.2 below).  As far as we are aware, these are the first determinations of the flaring
fraction of M dwarfs (and active M dwarfs) as a Galactic population.

\begin{figure}[htbp]
\centering
     \includegraphics[height=0.27\textheight]{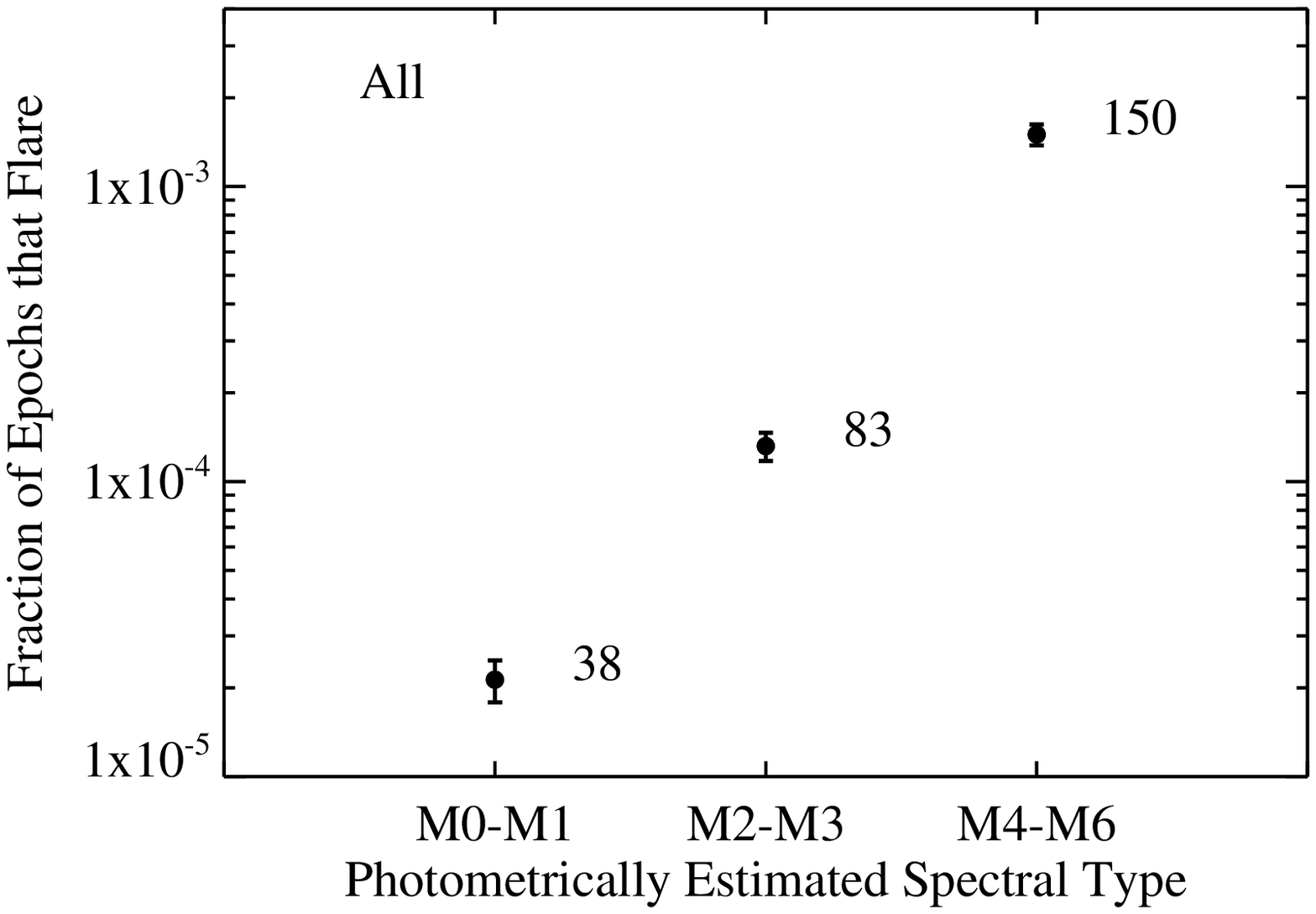}
     \includegraphics[height=0.27\textheight]{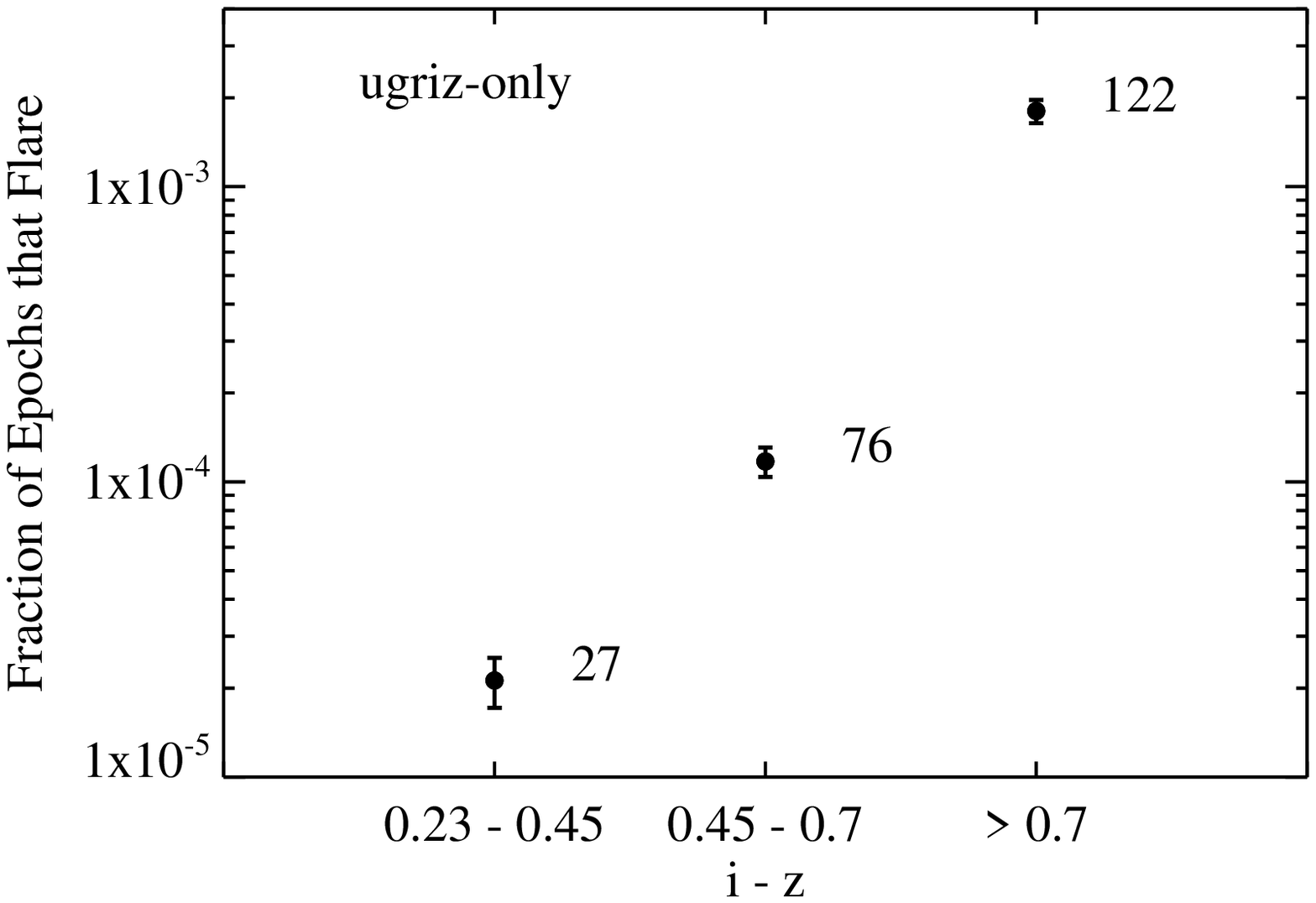}
      \includegraphics[height=0.27\textheight]{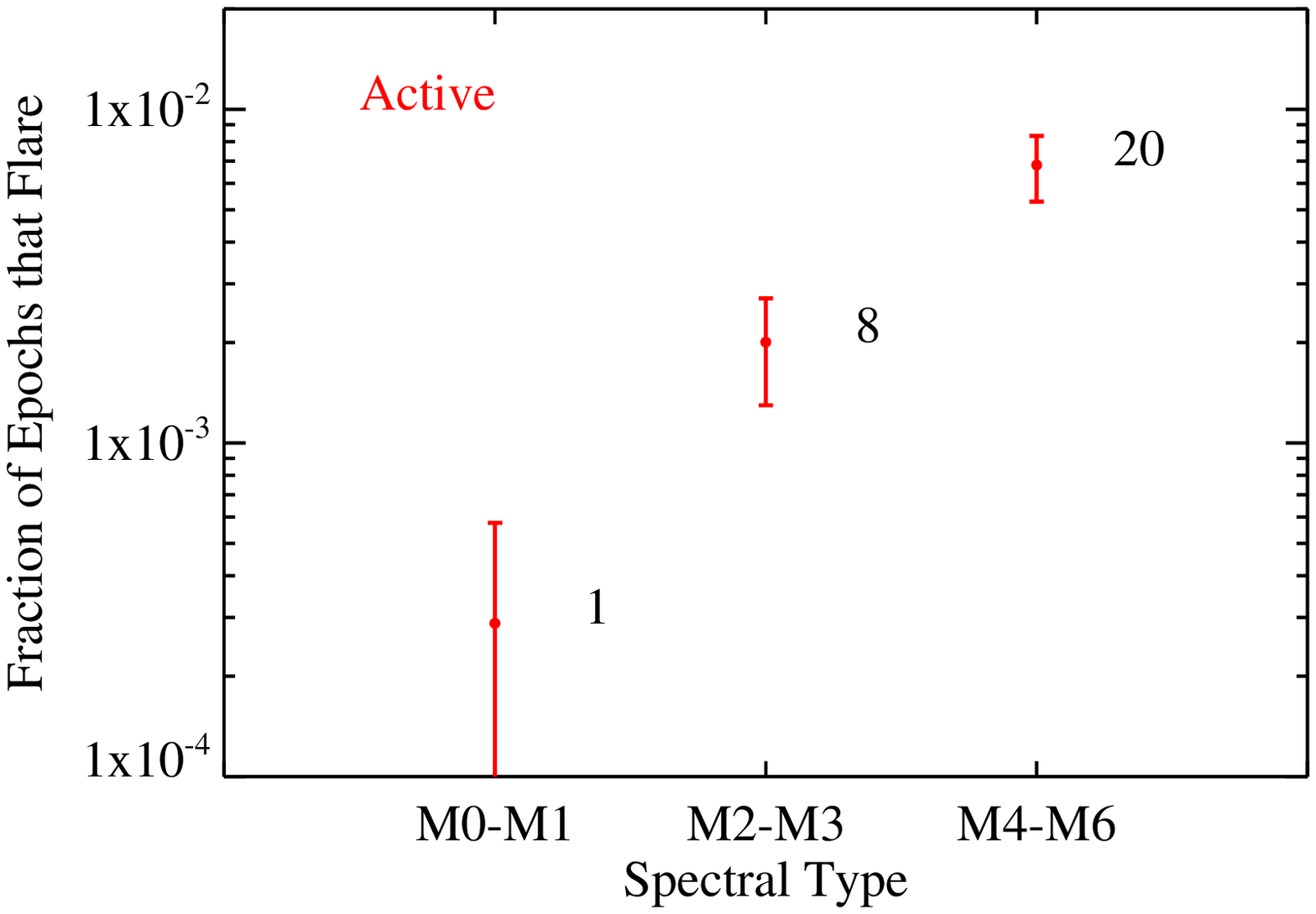}
      \includegraphics[height=0.27\textheight]{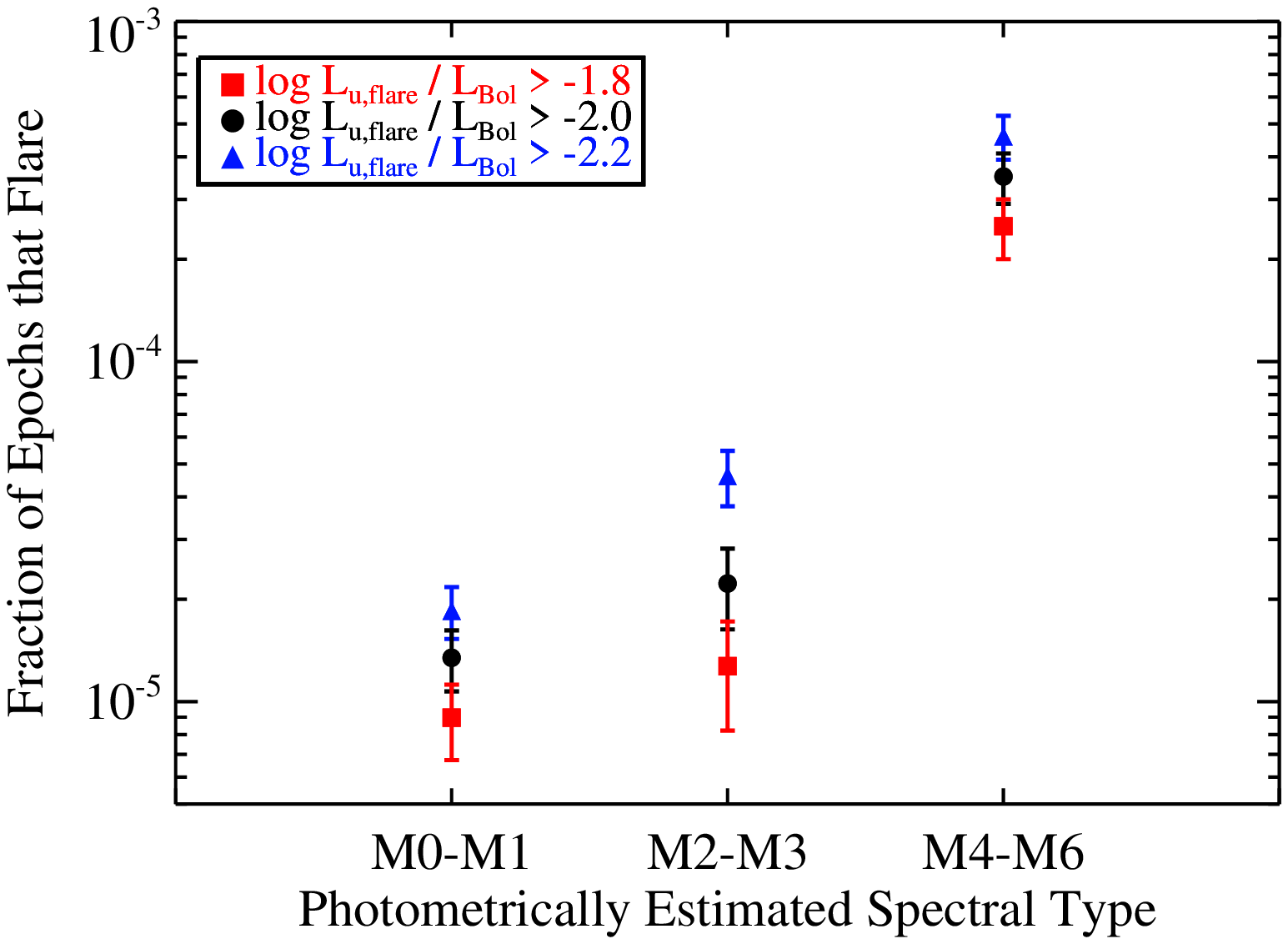}
    \caption{Top, left - The flaring fraction for the entire sample, binned according to the
photometrically estimated spectral type. The numbers indicate the number of flares in each 
spectral type bin. Top, right - The flaring fraction for the $ugriz$-only sample using $i - z$ color bins, showing
that using either color or spectral type isolates the same groups of stars.  Bottom, left - The flaring fraction for the active
stars with SDSS spectra only, binned according to the spectral type 
determined from each spectrum.  The flaring fractions
for the active stars are significantly larger than for the entire sample. Bottom, right - The fraction of 
epochs that are large flares, using several values of L$_{u,flare}$ / L$_{Bol}$ to define a large 
flare. Although it is easier to see low-luminosity flares in the late-type dwarfs (contrast effect), the
high-luminosity flares can be detected at all spectral types.  The number of flares in the (early, mid, late) type bins are (33, 29, 46), (24, 14, 35), and
(16, 8, 25) for the triangles, circles, and squares, respectively.}
\label{fig:flarefrac}
\end{figure}

Redder stars (later spectral types) also were found to flare more
frequently than earlier type stars, as expected because they are
more likely to be active and because lower energy flares are detectable 
due to their lower surface temperatures and fainter quiescent fluxes.
The top left panel of Figure \ref{fig:flarefrac} shows the flaring fraction
separated by spectral type group.  The fractions differ by about
an order of magnitude for each group, ranging from $2 \times 10^{-5}$ for the M0-M1 stars
to $1.3 \times 10^{-4}$ for the M2-M3 stars to $1.50 \times 10^{-3}$ for the M4-M6 stars.  It is notable
that the M4-M6 stars had a factor of 18 fewer observations (100,000
compared to 1.8 million) yet had nearly 4 times as many flares.  Clearly
the later-type stars form the bulk of the flare sample.  

Figure \ref{fig:hist_all} shows
that the total sample is sparsely populated by stars with spectral types M5 and later since they must 
be nearby to pass the $u < 22$ threshold.  In fact, $\sim$96\% of the observations in the M4-M6 bin are of M4 stars.  If we examine the M4 and M5-M6 bins separately, the flaring fractions become $1.2\times 10^{-3} \pm  1\times 10^{-4}$ for M4's and 
 $8.5\times 10^{-3} \pm  1.5\times 10^{-3}$ for M5-M6's, continuing the order-of-magnitude increase in flaring fraction towards later-types. However, the small number of stars (82), epochs (3656), and flares (31) on M5-M6 stars
in the sample makes this value relatively uncertain, so we don't include it in Figure \ref{fig:flarefrac}.

If we remove the dependence on activity by plotting
the flaring fraction of only the known active stars
 for each spectral type group (bottom left panel of Figure \ref{fig:flarefrac}), the fractions
are significantly higher but
the trend with spectral type remains.  The detection limit, $\Delta u \ge 0.7$, may be a contributing factor to the high flare fraction on the later spectral type stars. We can investigate this effect by considering only large flares, where a large flare is defined as having a $u$-band luminosity greater than 1\% of the bolometric luminosity (see \S 4.4 for details on the calculation of flare luminosities).  These higher luminosity flares will be detected at all spectral types in our sample.  Although there are fewer flares that are large (73), they provide a comparison of the flare fractions between early-, mid-, and late-types without a threshold bias.
Figure \ref{fig:flarefrac} (bottom right panel) shows the fraction of flaring epochs with large flares.  The flaring fraction for the later-type M dwarfs is still an order of magnitude larger than for the early-types, indicating that the detection threshold is not the
main factor contributing to this result.


\subsection{The Flaring of (Relatively) Inactive Stars}

In the SDSS spectroscopic sample, every star that flared showed
H$\alpha$ emission in its quiescent spectrum. 
To further investigate the activity status of the stars that flared but had no SDSS
spectra, we obtained followup spectra of 76 additional stars using the 
DIS spectrograph on the ARC 3.5m Telescope (see \S 2.3).  These spectra were spectral typed and 
had H$\alpha$ emission
measured using the Hammer analysis software \citep{Covey2007}.  The results are
given in Table \ref{table:DIS}.
Seven stars show no evidence of H$\alpha$ emission. One star, SDSS J022112.48+003225.7, has an ambiguous signature near 6563 \AA,
but this feature may be a blend of H$\alpha$ and a TiO band known to
exist in inactive M4 stars at this wavelength. Without higher resolution spectra,
we group this star among the inactive stars.  Another star, SDSS J204452.77$+$010546.9, exhibited 
variable H$\alpha$ emission between \emph{no} activity and \emph{weak} activity on
two different epochs (UT081107 \& UT080919).  Perhaps we happened to catch the 
UT080919 spectrum at the tail end of a flare; however, an analysis of the low-signal 
B400 spectra (3400 \AA$ - $5500 \AA)
showed no significant change in color between the two epochs, indicating that the observation
would have likely not met our flaring criteria.  We classify this object as \emph{variable},
a known trait of active M dwarfs \citep{Gizis2002}.

Of the eight likely inactive stars, four 
were observed on two separate nights to confirm the lack of emission.  
The spectral types for the eight stars show no obvious trend toward
early- or late-type (\textbf{Hammer type} (photometric type): \textbf{K7} (M0), \textbf{M0} (M0), 
\textbf{M0} (M0), \textbf{M1} (M0), \textbf{M1} (M0), \textbf{M3} (M3), \textbf{M4} (M4), \textbf{M4} (M3))
but the numbers are too small to draw any conclusion about the
spectral type distribution of inactive stars that flare.
Though formally classified as inactive, these stars may be
part of a group of weakly active M dwarfs which show Ca II H and K emission but
H$\alpha$ absorption \citep{Walk09}.  Certainly the fact that they
flared is a strong indication that they possess surface magnetic fields.  
 
From our combined spectroscopic sample (SDSS + DIS) of
(23+76 = 99) stars, with 8 being classified as inactive due to lack of H$\alpha$ emission, 
we derive a preliminary fraction of flaring stars that are inactive to be $\sim$8\% (the binomial errors
on this percentage are $1.9\%, +3.6\%$).  In other words, out of every 12 stars
that flare in the SDSS Stripe 82 region, one of them might be inactive.  This is an important value to
determine for modeling the rate of flares in the Galaxy, since as noted
in \S 2.2, most M dwarfs are inactive, especially at early-mid spectral
types.

\subsection{The Flaring Fraction vs. Distance from the Galactic Plane}

As a result of dynamical encounters, stars can move farther from the Galactic plane as they age.  The mean stellar age thus tends
to increase as a function of vertical distance $|Z|$ from the plane.
Therefore, other stellar properties that change with the lifetime of the star
should also change with $|Z|$ distance.  The fraction of active stars in the SDSS DR5 sample 
was found to decrease with $|Z|$ for each M spectral type \citep{West2008}, indicating that
magnetic activity depends on age. Many other studies have linked chromospheric activity to stellar age \citep{WW1970, Wielen1977, Giampapa1986, Soderblom1991, Hawley1996, Hawley1999, Hawley2000}.

\begin{figure}[htbp]
\centering
   \includegraphics[width=0.7\textwidth]{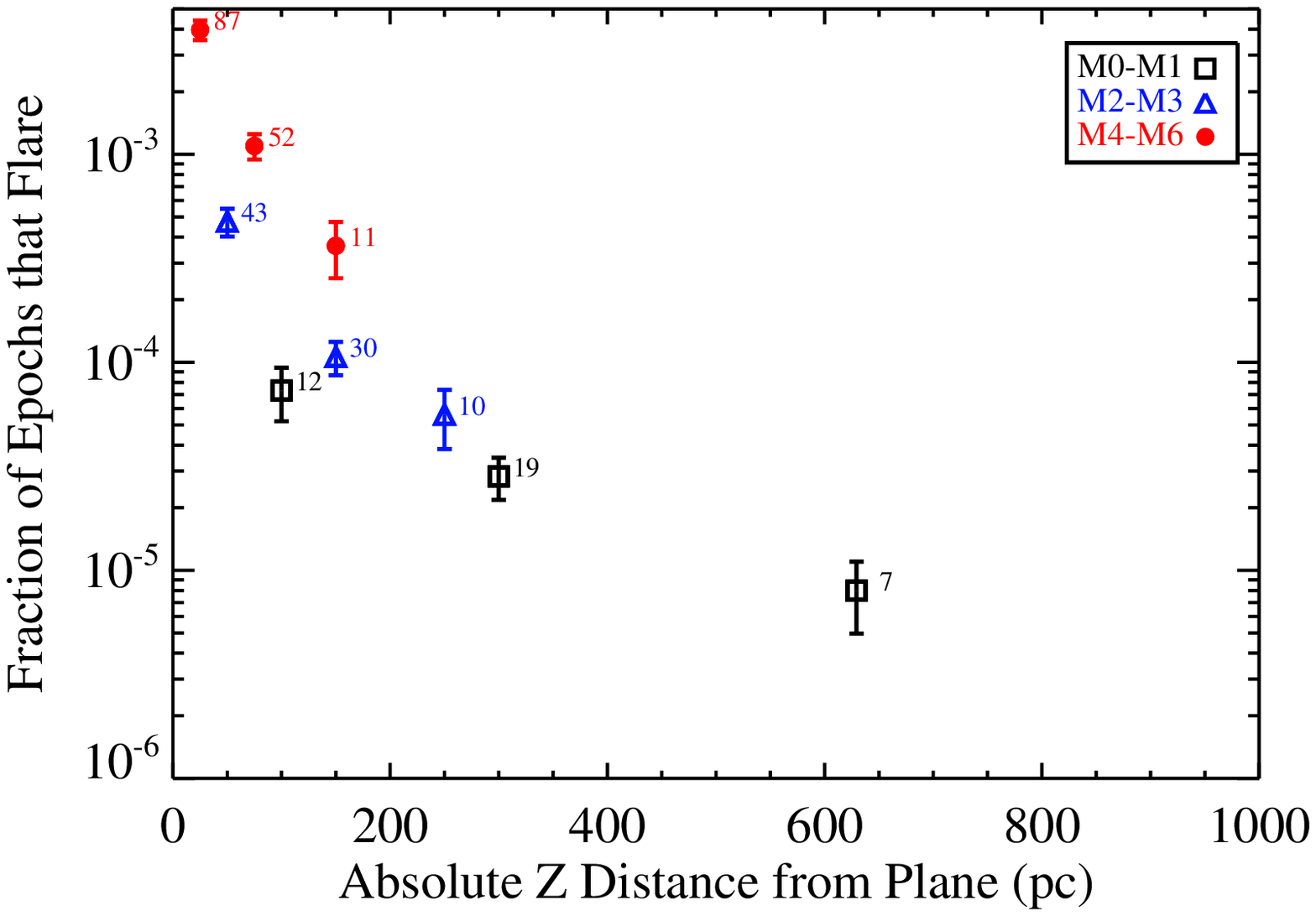}
   \includegraphics[width=0.7\textwidth]{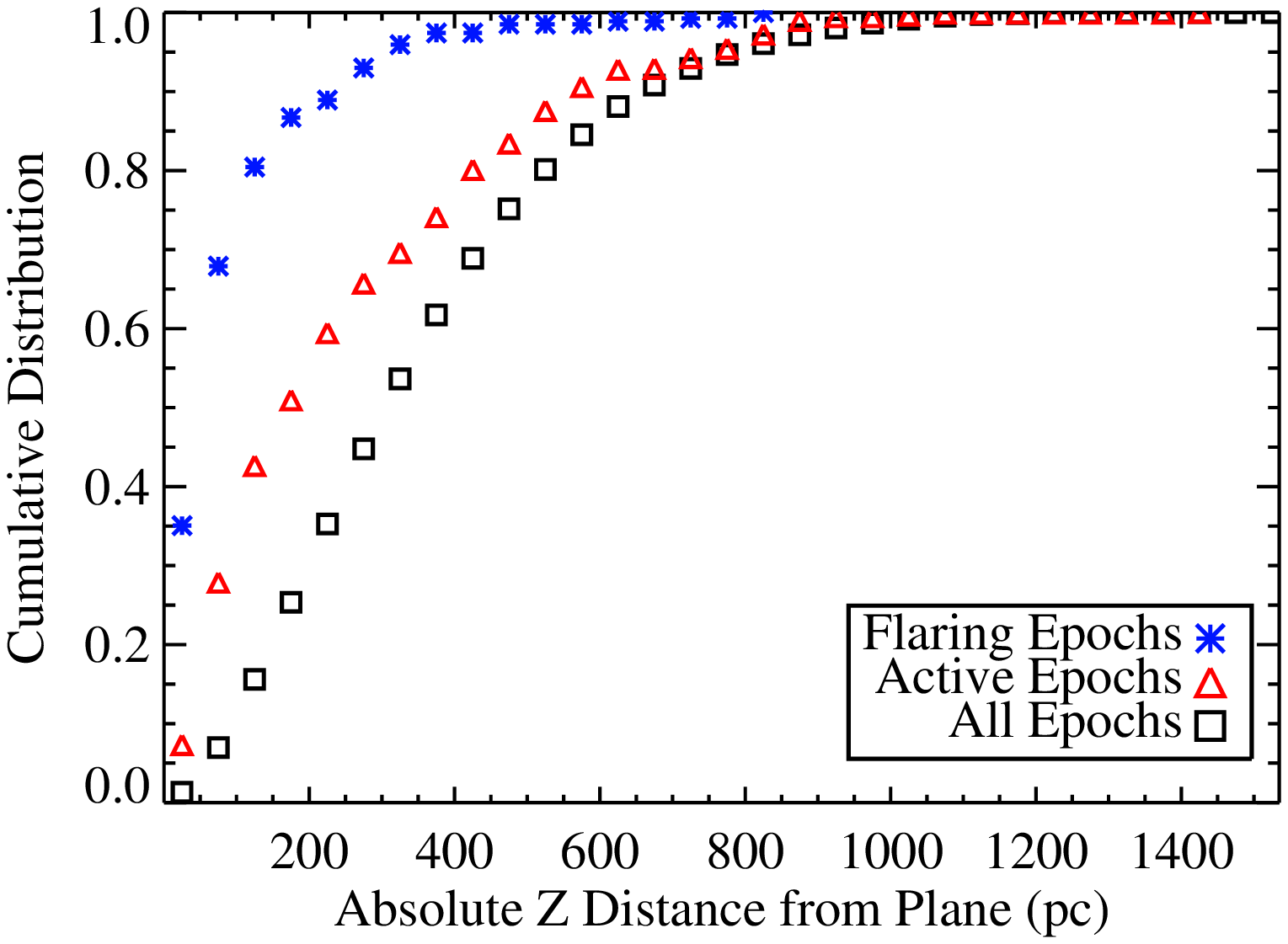}
 
   \caption{Top - The 
flaring fraction for different spectral type bins as a function of $|Z|$ distance from the plane.
The M0-M1 stars are plotted in bins of 0-200 pc, 200 - 400 pc, and 
400 - 900 pc; M -M3 stars
are plotted in bins of 0 - 100 pc, 100 - 200 pc, and 200 - 300 pc;  
M4-M6 stars are plotted in bins of 0 - 50 pc, 50 - 100 pc, and 
100 - 200 pc.  Note the differences in volume sampled for the different 
spectral types (due to the $u$ apparent magnitude limit).
Also, note the sharp decrease in flaring fraction as older populations 
are sampled farther from the Galactic plane.  Bottom -  The
cumulative distributions of $|Z|$ distance for all epochs (black squares), 
active epochs (red triangles), and flare epochs (blue asterisks), 
in bins of 50 pc.  The most distant flare occurs on a star that is
$\sim$850 pc below the plane, while
the most distant star is located at $\sim$1500 pc below the plane. }
\label{fig:Zdist}
\end{figure}

Figure \ref{fig:Zdist} shows the dependence of the flaring fraction on vertical
distance (age) for our three spectral type groups.  Computing the \emph{fraction} of 
epochs that flare effectively normalizes by the local stellar density and accounts for varying incompleteness 
in different distance bins.  Distances were
estimated using
an $r - z$ photometric
parallax relation \citep[in prep]{Boo09} for stars in the $ugriz$ sample and a
$g - K$ photometric parallax relation \citep{Covey2008champ} for stars 
in the $ug$-only
sample. As discussed previously, to obtain accurate estimates for 
the distances we applied
interstellar extinction corrections to the $r - z$ and $g - K$ colors before using
the photometric parallax relations.
The Sun was taken to be 15 pc above the Galactic plane \citep{Cohen1995, Ng1997, Binney1997}.
 The fraction of epochs where flares were detected decreases sharply with $|Z|$ distance for all three spectral
type groups.  This is expected  
because of the age dependence of (quiescent) magnetic activity, and the strong correlation
between active stars and stars that flare.  
As mentioned in \S 2.2, the volume over which the stars are measured differs
markedly for the different groups.  The later-type group 
is sampled only at close distances, while the mid- and early- type groups
are sampled progressively farther from the plane.
The cumulative distributions for the flaring epochs (asterisks), active epochs (triangles), and all epochs (open squares) 
are given in the bottom panel of Figure \ref{fig:Zdist} and
show the combined effects of age, spectral type and stellar density distribution
in an apparent-magnitude limited sample.  More than half the stars in the sample
are at a distance greater than 300 pc from the plane, but less than 10\% of the
flares occur on stars that far away.  Since the thin disk outnumbers the thick disk until $\sim$1 kpc \citep{Juric2008}, it is likely that these distant stars that flare still belong to the thin disk population.  Although 70\% of the active epochs are
within 300 pc, a third of the active stars are early-type stars which do not 
contribute significantly to the flaring population.  For these reasons and because of higher 
quality photometry for stars with $u < 21$, the nearby late-type, active stars clearly
make the largest contribution to the flaring fraction in this sample.  

The flaring fraction shows a larger decrease with increasing $|Z|$ distance than the active fraction from \cite{West2008}, especially
for the M4-M6 bin.  We find that an additional factor contributing to the trends seen in Figure \ref{fig:Zdist} is a bias towards
higher luminosity flares on more distant stars.  M dwarfs are known to exhibit a large spread in $u - g$ color \citep{Covey2008champ}.  For a given spectral type, we find a large spread in
$u$-band quiescent luminosity (see \S 4.4 for details on the calculation of flare luminosities).  For example, the M4's in this sample have an average $u$-band luminosity during quiescence of
$\sim$4$\times 10^{28}$ erg s$^{-1}$ with a spread of $\sim$2$\times 10^{28}$ erg s$^{-1}$.  For distances $ > 200$ pc,
most M4's become too faint in the $u$-band to be included in our sample.  At these large distances, only the M4's which are
intrinsically bright are included (i.e., there is a Malmquist bias due to our apparent
magnitude selected sample).  Therefore, a flare with $\Delta u = 0.7$ magnitudes on an intrinsically bright M4 has a larger $u$-band luminosity than a flare with the same apparent enhancement on an intrinsically fainter M4.  More luminous flares (or more luminous stages of flares) occur less frequently \citep{LME1976}, which generates
a bias in our sample towards smaller flare fractions at greater distances.  A useful future study would be to examine flare rates for a
flare-luminosity limited sample.

We also note that \cite{Welsh2007} found that most of the M dwarf flares detected in NUV Galex observations (1750 \AA$ - $2750 \AA) occur on stars within 300 pc of the plane; the flare flux in the NUV- and $u$-bands is thought to come from the same (or similar) emission mechanism, so it is expected that flare detections in these wavelength regimes
 should trace one another through the Galaxy.

\subsection{Flare Magnitudes and Luminosities}

\begin{figure}[htbp]
\centering
    \includegraphics[height=0.4\textheight]{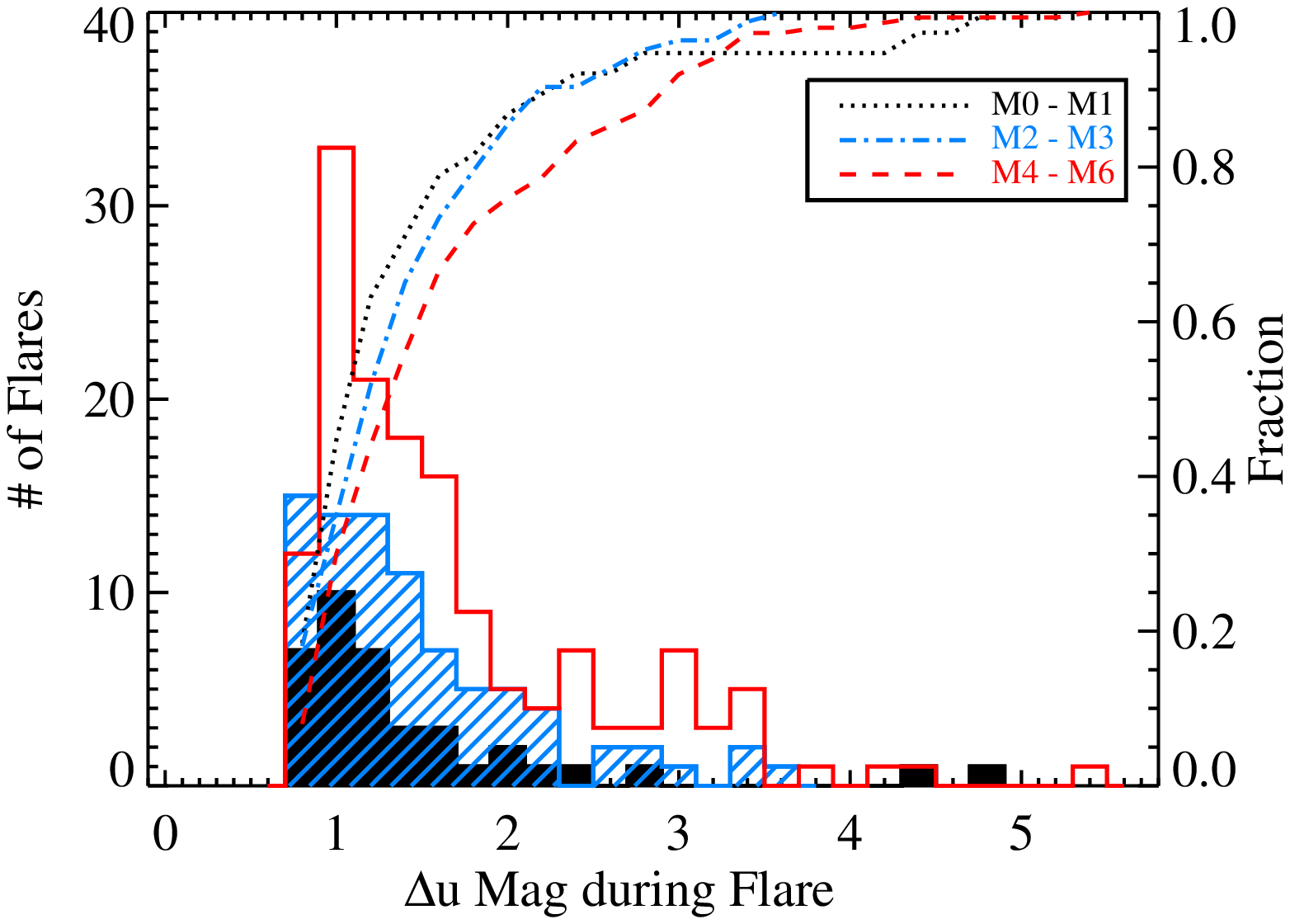}
    \includegraphics[height=0.4\textheight] {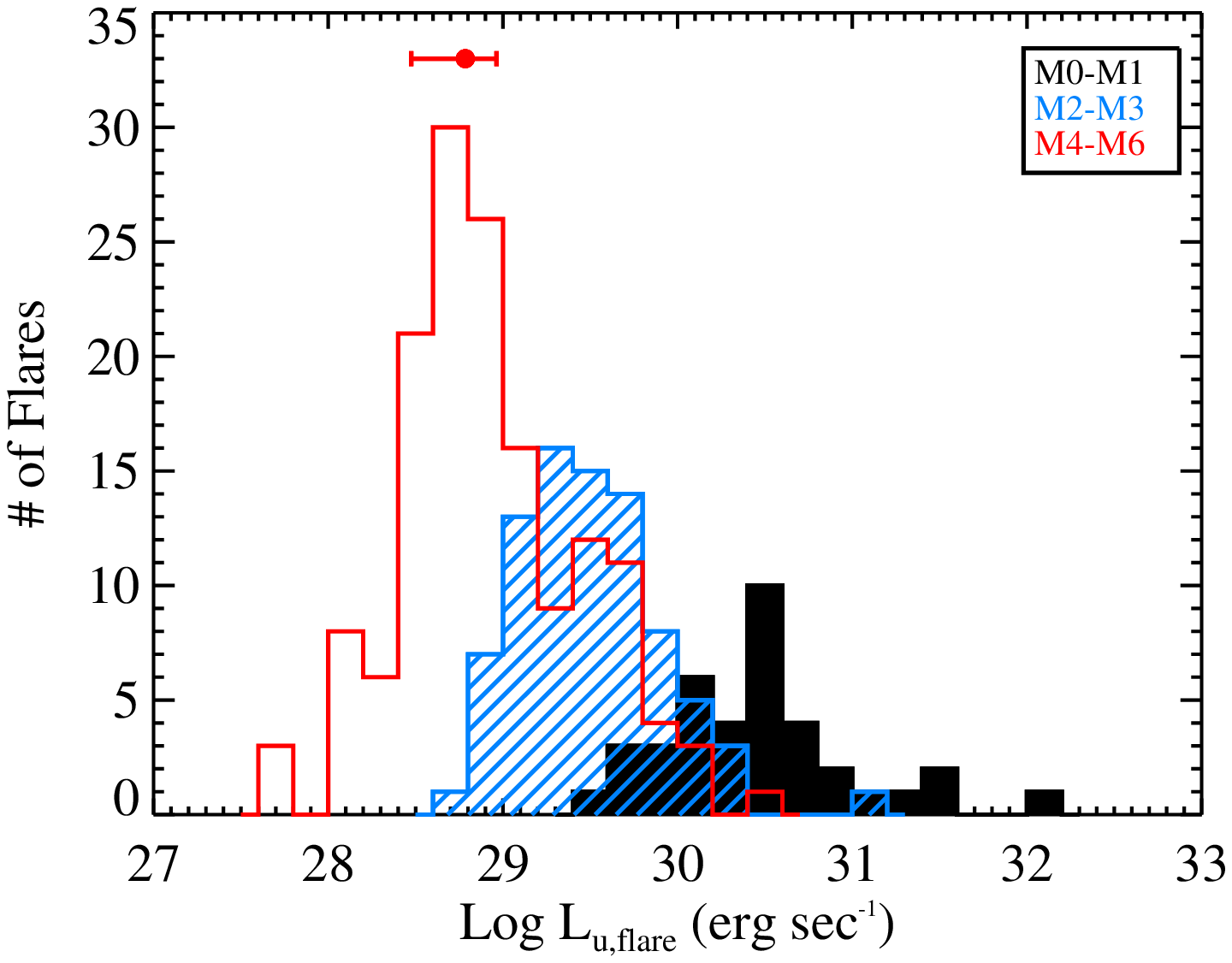}
    \caption{Top - The distribution of $u$-magnitude enhancements during 
flares.  The 
dotted lines show the cumulative distributions.  Later-type (lower mass) stars have the 
most detected flares and generate very large magnitude enhancements (due to their
smaller quiescent luminosity).  When converted to luminosities (bottom), the flares
on the higher mass stars are the most luminous.  Both of these results are consistent with
previous flare observations.  The errors in luminosities are $\sim$50\%, and we show a typical
error bar. }
\label{fig:flare_properties}
\end{figure}

Figure \ref{fig:flare_properties}, top, 
shows the distribution of flare magnitude ($\Delta u$) for the three
spectral type groups.  The later-type stars generally show larger magnitude
increases, due to their lower quiescent luminosity.  In other words,
it is much easier to see an energetically small flare on a star that is
intrinsically fainter.  Since the flare detection limit was chosen
to be $\Delta u \ge$ 0.7 mags on all stars, this corresponds to a lower luminosity
flare on the intrinsically fainter star. This relationship is consistent with previous 
flare studies \citep[e.g.,][]{Moffett1974, Rockenfeller}.  Figure \ref{fig:flare_properties} 
shows empirically which stars are likely
be observed with large magnitude increases, which is the metric of interest
for time domain surveys searching for transient brightness variations.

To obtain flare luminosities,
the quiescent stellar $u$-band flux was subtracted from each flare
observation, and the fluxes (which are per Hz) were converted to
luminosities using the distance
to each star (see \S 4.3) and the nominal $u$ bandwidth (FWHM 1.33$\times10^{14}$ Hz, or roughly $600$ \AA). Again, 
we use extincted-corrected magnitudes for the luminosity calculations. 
The flare luminosity distributions by spectral type are illustrated 
in the bottom panel of Figure \ref{fig:flare_properties}.  We calculate the
errors on the luminosities to be $\sim$$50$\%, which is dominated by the 
error in the photometric distance determination and includes the uncertainty in
the amount of red leak flux (up to 30\%) contained in the measured quiet $u$ magnitude.  
Systematic errors in the luminosity calculation are introduced since we assume full line-of-sight corrections for interstellar extinction.  As mentioned in \S2.2,  corrections effectively increase the distances, thereby increasing the 
luminosities by $\sim$10-70\%.

In Figure \ref{fig:flare_properties}, the low luminosity cutoff of each distribution represents the detection limit
of $\Delta u = 0.7$ mags, which corresponds to nearly an order of 
magnitude difference in 
luminosity for each group
(log L $\sim$ 28.0 for M4-6, $\sim$ 28.7 for M2-3, $\sim$ 29.5 for M0-1).
Although the flare magnitude enhancements on the earlier-type stars are smaller (as
shown in the top panel),
their derived luminosities are generally larger than the flares on the later-type stars.
This is consistent with the findings of \cite{LME1976} and \cite{Pettersen1988}, that 
the average rate of energy loss in the $U$-band during flares increases as a function of stellar bolometric luminosity.  For example, if
we compare the two largest $u$-magnitude enhancements in the SDSS flares, 
the $\Delta$$u = 4.7$ flare on a M1 star gives a luminosity of 
1.3 $\times$ 10$^{32}$ erg s$^{-1}$ while the 
$\Delta$$u = 5.5$ flare on a M6 star gives a luminosity
 of only  1.7 $\times$ 10$^{29}$ erg s$^{-1}$.
However, we emphasize that our derived flare luminosity is 
sampled at a random time during the flare
and there is no way to determine whether the emission occurred during the rise,
peak or decay phase so it is not strictly comparable to the average energy.
It is interesting that we do not observe very high
luminosity ($> 10^{30}$ erg s$^{-1}$) flares on the very red stars.  This is likely due
to two factors: they are quite rare on the relatively low (quiescent) luminosity stars; and
the number of epochs that SDSS observed on the later-type stars is much smaller than on
the earlier type stars since the latter comprise a much larger fraction of the sample.
When we compare the flare luminosity to the bolometric luminosity of 
the stars \citep{NLDS}, we find a few flares that emit as much as $\sim$1-10\% of the bolometric luminosity in the $u$-band, on stars of 
both early and late spectral types.  \cite{Pettersen1988} had determined 
a limit of $10^{-4}$ for the average L$_{U}$ / L$_{bol}$; however, our values may represent instantaneous
measurements of more luminous stages during flares.  

With a common flare detection threshold ($\Delta u = 0.7$ magnitudes) for all spectral types, the late-type stars emit
less luminous, yet many more, flares.  However, we find that the \emph{total flare energy} emitted
on early-, mid-, and late-type stars is approximately equal for equal observing times.  For example,
the peak luminosities of Figure \ref{fig:flare_properties} occur at log L $\sim$ 30.5, 29.3, and 28.7 erg s$^{-1}$, a ratio of $\sim$ 60 : 4 : 1 
for early-, mid-, and late-types, respectively.  When we sum the luminosities, multiply by the filter integration time,
and scale to the same total number of observations (\S 3.4), the total flare energies occur in the ratio $\sim$ 0.7 : 0.3 : 1.  
Apparently, the larger flare rate of the late-types approximately makes up for the lower energy output per flare, suggesting
that early- and late-type stars, \emph{summed over a range of quiescent activity levels and stellar ages}, may possess similar amounts of total stored magnetic energy that is eventually released as flares, but that this flare energy is released at different rates on stars of different masses.  Further study of a sample including smaller flares is needed to confirm this preliminary result.

\subsection{Flaring Frequency}
\begin{figure}[htbp]
\centering
    \includegraphics[width=0.9\textwidth]{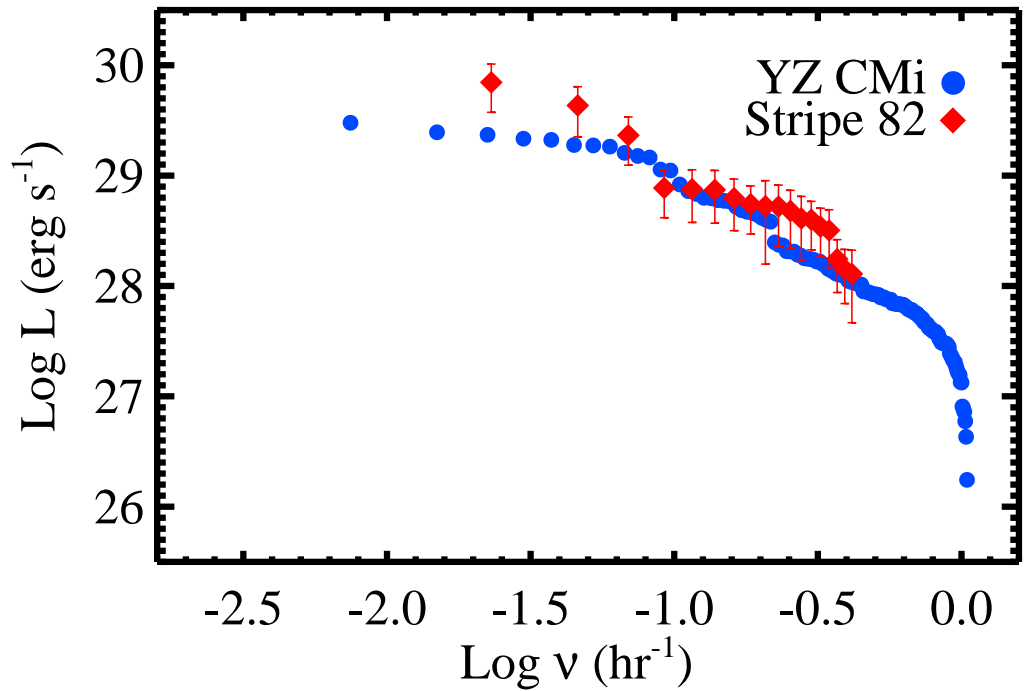}
    \caption{The flaring frequency for active M4-M5 stars from Stripe 82 
compared to the average $U$-band flare frequency of the 
very active, nearby dM4.5e star, YZ CMi \citep{LME1976}.  The good
agreement between the two distributions demonstrates that flaring frequencies obtained
from observing a large number of stars at a low cadence is consistent with 
observing a single flare star continuously.  The SDSS Stripe 82 
distribution contains 18 flares.}

\label{fig:LME}
\end{figure}

  A classic study of flare frequency for a handful of nearby,
very active M dwarfs was carried out by \cite{LME1976}.  They used
a monumental collection of continuous
photometric monitoring data \citep{Moffett1974}, which provided information throughout the
time evolution 
of each flare.  An average luminosity for each flare may be calculated
by dividing the reported total $U$-band energy of the flare by its duration. In 
Fig. \ref{fig:LME}, we compare
the luminosities  of the SDSS flares (sampled at a random time during the flare)
to the average flare luminosities of the Moffett flares for the dM4.5e star YZ CMi.  
In order to estimate the frequency with which a flare of a particular 
luminosity (or greater) was
seen, we used the total $u$-band observing time
for active stars with SDSS spectra and spectral types of M4 or M5 
(43.335  hours), and obtained a cumulative distribution of 
flare luminosities.  As shown in Figure \ref{fig:LME},
our derived flaring frequency from the SDSS data is remarkably consistent
with the flaring frequency of YZ CMi, following a power law form with more luminous 
flares occurring less frequently.

\begin{figure}[htbp]
\centering
    \includegraphics[width=0.9\textwidth]{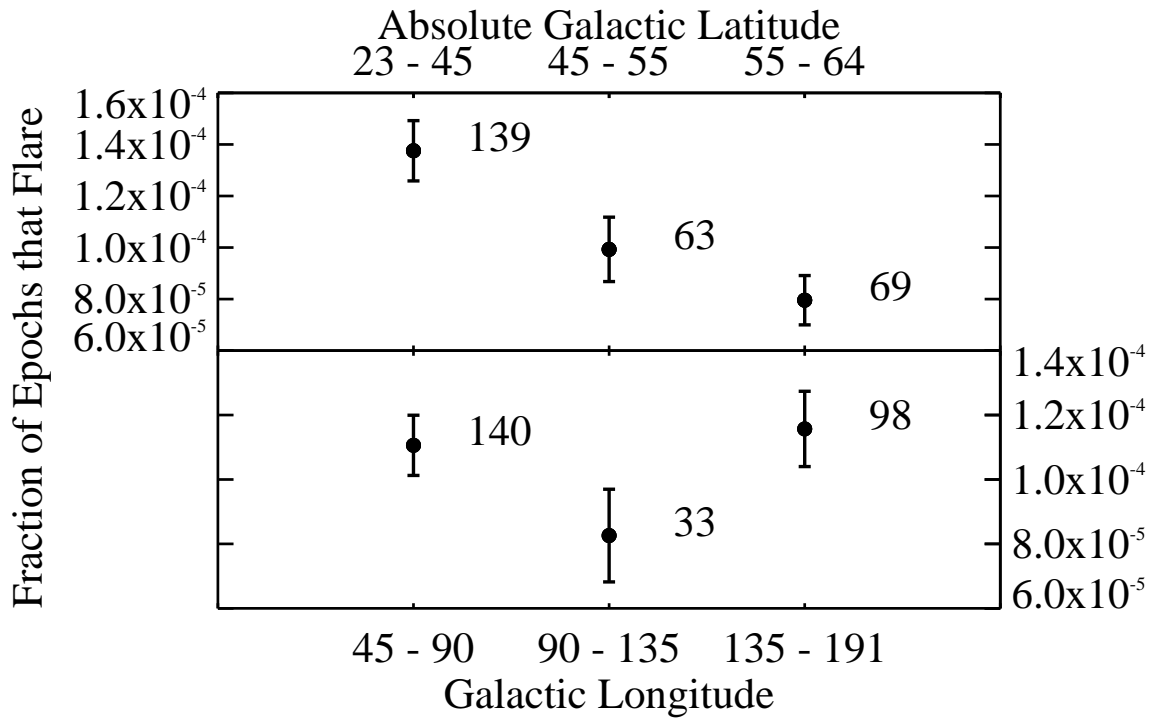}
    \caption{The flaring fraction for 3 divisions of
Galactic latitude (top) and longitude (bottom).  The flaring fraction 
decreases for lines of sight with larger Galactic latitude, probably because a smaller 
fraction of the sample is close to the Galactic plane (and therefore likely to be younger and magnetically active).  There
is no obvious trend with Galactic longitude.  Note that the flaring 
fraction is a normalized quantity, so number density variations have been removed.  }

\label{fig:latlong}
\end{figure}

\begin{subsection}{Spatial Flare Rate in Stripe 82}

A number of interest to characterize transient events in
large sky surveys is the spatial flare rate, the number of flares per hour
per square degree.  Of course this rate will vary depending on
which area of the sky is being observed (e.g. with Galactic latitude).
For the Stripe 82 data, which span \emph{l} = (45$\degr$ to 191$\degr$); \emph{b} = (-23$\degr$ to -64$\degr$) the rate
can be empirically calculated by
taking the number of observed flares (271), and dividing by
the spatial extent of the Stripe 82 time domain database 
(279.7 square degrees) and the average
observing time (50 epochs x 54 sec per epoch = 2700 sec).
\emph{This gives a rate of 1.3 flares hr$^{-1}$ sq deg$^{-1}$ with a $u$ mag increase of at least 0.7 mags.}

The derived spatial flare rate is a lower limit to the intrinsic flare rate because we have an inefficiency, $\epsilon$, for detecting flares.  Major contributions
to $\epsilon$ are varying contrast levels for the different
spectral types, the fact that not all flares produce an increase 
in the $u$ and $g$ bands (e.g., flares that begin after the $u$ but before the $g$ observation), our 
limit of 10\% false positives by using a $\Phi_{ug}$ threshold of 100, and our
strict $u < 22$ magnitude cutoff.  As noted explicitly, we intentionally adopted
a conservative $\Delta u = 0.7$ magnitude limit for flare detection, so our spatial
flare rate is only representative of relatively bright flares.  

We calculated the number of flares divided by the average observing time per star ($\sim$2700 sec) for every square degree (in RA and DEC) in Stripe 82 to investigate how 
the number of flares hr$^{-1}$ sq deg$^{-1}$ varies with the line-of-sight through the Galaxy.  We find a maximum flare rate of 8 flares
hr$^{-1}$ sq deg$^{-1}$ occurring closer to the plane ($l \sim 56\degr, b \sim -38\degr$) where the density of stars is higher
 ($\sim$240 stars sq deg$^{-1}$ with $u < 22$ compared to the average for Stripe 82,
$\sim$180 stars sq deg$^{-1}$ with $u < 22$) and where a higher fraction of stars are active (since they
are closer to the plane and therefore likely to be younger).  

To illustrate how the spatial flare rate varies with Galactic latitude and longitude, 
we calculated the average flare rate
for 3 latitude regions and 3 longitude regions.\footnote{When binning Stripe 82 into $(l, b)$, we divided the flare rate in each 1 deg
 $\times$ 1 deg square by cos($b$) in order to correct for the 
changing area due to converging lines of longitude.  To compensate for incomplete coverage at the edges of Stripe 82,
we also divide by the fraction of each 1 deg
 $\times$ 1 deg bin that was scanned by Stripe 82, and we consider only bins that have $> $50\% coverage.}
Our results are presented in Table \ref{table:spatialdivisiontable}.  This raw spatial flare rate decreases with increasing Galactic latitude, as expected due to the decreasing
stellar density farther from the plane.  To account for stellar density, we divided by the number of total epochs in each region and present the
flaring fraction as a function of Galactic latitude and longitude in Figure \ref{fig:latlong}.  There is no obvious trend
with longitude; however, the flaring fraction generally decreases with increasing latitude, probably because lines-of-sight
at higher latitude include a smaller fraction of stars at small $|Z|$ distance (see \S 4.3).

Finally, the empirical flare rate for all M dwarfs would also include the flares occurring
on the M dwarf - white dwarf binary systems that we excluded in step 8 of \S 2.1.  We have
searched those light curves and find 6 events that would satisfy our flare criteria.
This has a negligible effect on the overall flare rate.

\begin{subsubsection}{Comparison to Simple Model Prediction}
For comparison, we can also estimate the flare rate 
from our knowledge of the spatial
density of M dwarfs in Stripe 82 (Figure 3), the fraction
of active stars at each spectral type (Figure 3), the
relative flare rates for active and inactive stars (\S 4.1, 4.2),
and the flare frequency distributions for stars of different
spectral types \citep{LME1976}.  This simple model calculation gives
a prediction of 
the spatial flare rate we should observe in Stripe 82.

We consider the same spectral type ranges, and the values
measured or derived previously to form the general equation:

\begin{equation}
 \mbox{\# flares} = s \times (f_{active} \times f_{a,flare} + f_{inactive} \times f_{i,flare} ) \times \nu \times t 
\end{equation}
where $s$ is the number of stars, $f_{active}$ is the fraction of stars 
that are active, $f_{a,flare}$ is the fraction of active stars that flare,
 $f_{inactive}$ is the fraction of stars 
that are inactive, $f_{i,flare}$ is the fraction of inactive stars that flare,
$\nu$ is the number of flares hr$^{-1}$ above a limiting luminosity, and
$t$ is the observing time per star.  A very uncertain parameter is $f_{i,flare}$,
and we calculate the predicted number of flares adopting both 1\% and 10\%.\footnote{In \S4.2 we found that 
$\sim$8\% of the flare stars are inactive.  Although this is not strictly the same as $f_{i, flare}$, which is the
fraction of inactive stars that flare, the 8\% value made the adopted range of 1 - 10\% seem reasonable.}  We list the
measured and derived values for Equation 4 in Table \ref{table:spatialtable}.




In \S 3.4 we found that the number of confirmed (observed) flares
was 38 for M0-M1 (compare to 20 - 66 predicted); 83 for M2-M3 (compare
to 68 - 150 predicted); and 150 for M4-M6 (193 - 233 predicted).  In all cases the predicted 
values are reasonably close to the observed number of flares.  Given
the crude estimates in our predictions, this agreement is encouraging,
indicating that our method for finding flares in the low-cadence SDSS
photometric data is apparently reliable.  

There are also several places where the predictions
can be improved.  Two very poorly determined parameters are the fraction of inactive stars that flare and
the flaring frequency, $\nu_{i}$, of inactive stars that flare
(here taken to be the same as the flaring frequency, $\nu$, of active stars of the same
spectral type).  We have an ongoing program to investigate the flaring
rates of inactive stars through continuous monitoring observations \citep[in prep]{Hilton09}.

Furthermore, our estimates for $\nu$ are taken from YZ CMi, a star with a 
well-determined but likely higher-than-average flare frequency distribution \citep{LME1976}.  Therefore, it is not surprising that 
our predicted number of flares for this bin exceeds the observed number.  Even if we
set $\nu_{i} = 0$ flares hr$^{-1}$, 189 flares are predicted. If YZ CMi accurately represents
 the flaring frequency for active M4-M6 stars, we estimate our inefficiency, $\epsilon$, for detecting flares above $\Delta u \ge 0.7$ mags on M4-M6 stars to be
$>  20$\%, assuming all the other parameters are correct. 
Alternatively, if instead we use a slightly lower flaring
frequency for all M4-M6 stars, $\nu = 0.3$ flares hr$^{-1}$, we predict $145 - 175$ flares, which is 
consistent with the observed value.


We also note that the active fraction of M4 stars found here (30\%)
is considerably lower than the active fraction found from the full
DR5 sample (50\%, West et al. 2008).  DR5 is dominated by stars
in the Northern Galactic cap, with most observations obtained 
looking out of the Galactic disk.  In contrast, Stripe 82 is
observed through the disk, looking toward the Southern Galactic cap.
The difference in active fraction between the North and South
is intriguing but unexplained.  We are pursuing an additional study
of the change in active fraction with galactic latitude.

Several regions in Stripe 82 have moderately high extinction ($ > 0.5$ mags in the
$g$-band), and we examined the possibility that the spatial flare rate is correlated 
with star-forming regions and therefore contaminated by accretion events.  We compared the 
composite H$\alpha$ map from \cite{Finkbeiner2003} to the spatial distribution of 
stars that flare and note an absence of any conspicuous 
clustering of flare stars near H$\alpha$ regions.  The star-forming regions MBM 18/MBM 19, found at RA $> 55\degr$, coincide
with the areas of greatest extinction in Stripe 82 and are suspected sites of low-mass star formation \citep{Peregrine}. 
None of the 11 flare stars with RA $> 55\degr$ have spectra to examine for accretion features.  Of the 
stars that do have spectra, none have especially high equivalent widths characteristic of accretors.  

\end{subsubsection}

\end{subsection}

\section{Conclusions}
  We used the multi-epoch $ugriz$ photometry of the SDSS Stripe 82 to characterize
the M dwarf flaring properties as a function of spectral type, level of magnetic activity, distance
from the Galactic plane, and line-of-sight through the Galaxy.  In contrast to studying flare rates on a few individual 
nearby stars via continuous monitoring, we quantify the flaring properties of \emph{populations} of M dwarfs using
low-cadence observations. 
We search for a flare signature on an epoch-by-epoch basis over millions of observations and employ
the FDR algorithm to limit the number of false positives.  We find 271 flares among $\sim$2.5 million
observations.  The fraction of epochs that flare increases by nearly an order
of magnitude for each subsequently redder M dwarf color bin.  We attribute this trend to there 
being better `flare visibility' and an increasing fraction of magnetically active stars
in the redder bins.

 We find a strong correlation between stars that have flares and stars that 
have H$\alpha$ in emission during quiescence. For the sample as a whole, 
1 out of 10,000 observations shows a flare, but $\sim$30 out of 10,000 
epochs observed on active stars show flares.  This finding is consistent 
with a simple physical picture
connecting magnetic activity to flaring: surface magnetic 
loops (the source of the persistent H$\alpha$ emission line) reconnect and 
accelerate charged particles, which then deliver energy to the lower 
atmosphere and produce the optical flare emission.  However,
the detailed physics of how such large energies in the $u$-band are 
generated remains a mystery.  Unfortunately, the time difference between 
the $u$ and $g$ filters adds ambiguity to the interpretation of the flare 
emission spectrum and prevents us from constraining the broadband spectral 
shape of flares with these data.
 
 Our total sample is dominated by inactive (no H$\alpha$ emission) M dwarfs, 
and we find only eight flaring stars that are classified as inactive.  Inactive stars
may have a much smaller filling factor of magnetic loops such that they do not show quiescent
activity but still have a (small) chance of producing flares.
Continuous monitoring of inactive M dwarfs is needed to more accurately 
quantify the flaring rates and properties of these stars.

  The amplitudes and luminosities of the SDSS flares are consistent 
with previous 
observations.  Most flares (80\%) have amplitudes less than 2 magnitudes 
in $u$, which is not surprising since random imaging is most 
likely to catch a flare during its more extended, smaller-amplitude
decay phase.  Both early- and late-type stars exhibit $u$-band magnitude 
enhancements as high as $\Delta$$u$ $\sim$ 5; however, the actual 
luminosity for a given magnitude enhancement is dependent on the 
intrinsic shape of the underlying stellar spectrum.  Very large 
luminosity flares are not seen on the reddest of stars, which may be a 
combination of a real physical effect and a result simply from there not 
being as many observations of very red stars in the sample.  The most 
luminous flares in the sample emitted $\sim$10\% of the stellar 
bolometric luminosity at the time they were sampled.

  The flaring fraction is found to strongly decrease with increasing $|Z|$ 
distance for all spectral types, as expected, due to a smaller active fraction and a Malmquist bias at greater distances.   Our
best estimate of the lower limit on the flaring rate (averaged over Stripe 82) for flares with $\Delta u$ $\ge$ 0.7 on
stars with $u < 22$ is 1.3  flares hour$^{-1}$ square degree$^{-1}$, and the larger spatial flare rates and fractions
are typically found along lines-of-sight at lower Galactic latitude in Stripe 82.  Future time-domain surveys, such
as LSST, will be able to probe to a much larger distance from the plane; however, we predict that the spatial flare rate for flares of $\Delta u \ge 0.7$ mags on stars from M0 - M4 will not
increase appreciably for intermediate to high Galactic latitudes due 
to our observed $|Z|$ distance-flaring relation.  For lines-of-sight 
through the Galactic plane, there is likely to be a significant increase 
in the observed flare rate. Since the $u$-band will
be observed only infrequently with LSST, the 
$g$- and $r$-bands will be the most useful to identify flaring epochs.  
We find that $\sim$95\% of the flares have $g$ magnitude enhancements 
of $< 1$ magnitude.  However, due to the higher precision ($\sim$5
mmag at the bright end; Ivezi\'{c} et al. 2008) of LSST,
a larger number of small amplitude flares will be observed.  For these 
reasons, flares will likely form a \emph{larger} population of transients 
in LSST compared to SDSS, which was limited by
the photometric uncertainties in the $u$-band.  Since the SDSS flares 
were selected from the $u$- and $g$-bands, the properties of these flares 
in the other bands could be applied to candidate flares in future 
surveys which do not use the $u$-band.

\section{Acknowledgments}
 The authors would like to the thank the referee, Terry Oswalt, whose comments
greatly improved this paper.  We also thank Chris Miller and Andy Connolly for their insightful
discussions regarding the False Discovery Rate, and Andy Gould for 
suggesting the importance of followup observations of the stars that 
flared.  We are grateful to John Wisniewski for sharing APO 3.5m observing time with us.  AFK, SLH, and EJH acknowledge support from NSF grant
AST 08-07205.  JJB and SLH acknowledge support from NSF grant AST 06-07644.
 
   Funding for the SDSS and SDSS-II has been provided by the Alfred P. Sloan Foundation, the Participating Institutions, the National Science Foundation, the U.S. Department of Energy, the National Aeronautics and Space Administration, the Japanese Monbukagakusho, the Max Planck Society, and the Higher Education Funding Council for England. The SDSS Web Site is http://www.sdss.org/.

    The SDSS is managed by the Astrophysical Research Consortium for the Participating Institutions. The Participating Institutions are the American Museum of Natural History, Astrophysical Institute Potsdam, University of Basel, University of Cambridge, Case Western Reserve University, University of Chicago, Drexel University, Fermilab, the Institute for Advanced Study, the Japan Participation Group, Johns Hopkins University, the Joint Institute for Nuclear Astrophysics, the Kavli Institute for Particle Astrophysics and Cosmology, the Korean Scientist Group, the Chinese Academy of Sciences (LAMOST), Los Alamos National Laboratory, the Max-Planck-Institute for Astronomy (MPIA), the Max-Planck-Institute for Astrophysics (MPA), New Mexico State University, Ohio State University, University of Pittsburgh, University of Portsmouth, Princeton University, the United States Naval Observatory, and the University of Washington.

\clearpage
\clearpage

\clearpage
\clearpage

\clearpage
\clearpage

\clearpage
\clearpage

\clearpage
\clearpage

\clearpage

\vspace{25pt}

\clearpage

\clearpage

\clearpage
\clearpage

\clearpage
\clearpage

\clearpage
\clearpage

\clearpage

\clearpage

\clearpage
\begin{deluxetable}{ccc}
\tablecolumns{3}
\tabletypesize{\scriptsize}
\tablewidth{0pc}
\tablecaption{M Dwarf Colors and Spectral Types}
\tablehead{
\colhead{Spectral Subtype}    &  \colhead{$\mu$}   & \colhead{$\Sigma$}}

\startdata

M0 & $(0.55, 0.32)$ & $ \left(\begin{array}{cc}
0.0159 & 0.0039 \\
0.0039 & 0.0083 \end{array} \right)$ \\
M1 & $(0.78, 0.43)$ & $ \left(\begin{array}{cc}
0.0135 & 0.0051 \\
0.0051 & 0.0055 \end{array} \right)$ \\
M2 & $(0.95, 0.52)$ & $ \left(\begin{array}{cc}
0.0163 & 0.0061 \\
0.0061 & 0.0058 \end{array} \right)$  \\
M3 & $(1.15, 0.62)$ & $ \left(\begin{array}{cc}
0.0204 & 0.0076 \\
0.0076 & 0.0076 \end{array} \right)$  \\
M4 & $(1.41, 0.76)$ & $ \left(\begin{array}{cc}
0.0213 & 0.0086 \\
0.0086 & 0.0085 \end{array} \right)$  \\
M5 & $(1.87, 1.03)$ & $ \left(\begin{array}{cc}
0.0183 & 0.0084 \\
0.0084 & 0.0062 \end{array} \right)$  \\
M6 & $(2.09, 1.14)$ & $ \left(\begin{array}{cc}
0.0194 & 0.0072 \\
0.0072 & 0.0054 \end{array} \right)$  \\
M7 & $(2.49, 1.35)$ & $ \left(\begin{array}{cc}
0.0270 & 0.0107 \\
0.0107 & 0.0095 \end{array} \right)$  \\

\enddata
\tablecomments{The best fit parameters at each M dwarf spectral type for each 2D 
Gaussian distribution of $r - i$, $i - z$ colors, determined from 20,000
M dwarfs in the \citet {West2008} DR5 spectroscopic sample
with well-measured colors and well-determined types.  $\Sigma$ is the covariance
matrix and is of the form  $\Sigma =  \left(\begin{array}{ccc}
 \sigma_{r-i}^2 & \sigma_{r-i,i-z}  \\
 \sigma_{r-i,i-z} & \sigma_{i-z}^2 \end{array} \right)$.  The 
probability for each spectral type is evaluated as 
$p(x) = \frac{1}{2\pi \| \Sigma \| ^{1/2}} exp(-\frac{1}{2}(x - \mu)^T \Sigma^{-1}(x - \mu))$, 
$\mu =  \left(\begin{array}{cc} \mu_{r-i} \\ \mu_{i-z} \end{array} \right) $ are the mean colors for a given type, and $x$ are the 
$r - i$, $i - z$ colors for a given star.}
\label{table:photclass_param}
\end{deluxetable}
\clearpage

\clearpage
\begin{deluxetable}{llllll}
 
 \tabletypesize{\scriptsize}
 \tablecaption{Flare Selection Criteria}

 
\tablehead{
\colhead{$Criterium$}&
\colhead{$ugriz$}&
\colhead{$ug$}&
\colhead{$Spectroscopic (SDSS)$}&
\colhead{$Justification$}&
}
\startdata
\\
Clean Sample & 39,633 (1,987,253) &  10,497 (524,981) & 4,005 (197,867) & M star colors, no WD-dM \\
$\Phi_{ug}> 0$ & 39,629 (1,031,706) & 10,497 (272,249) & 4,004 (102,582) & $\Phi_{ug} < 0$ from noise in $u$ and $g$ \\ 
$F_{u,j}-F_{u,quiet} > 0,$ & 39,620 (612,728) & 10,484 (158,550) & 4,004 (61,107) & Eliminates possible eclipses \\ 
$F_{g,j}-F_{g,quiet} > 0$ \\
$\Phi_{ug} \ge 100$ & 321 (463) & 132 (210) & 36 (51) & FDR \\
$\Delta u$ mag $\ge 0.7$ & 243 (296) & 60 (74) & 29 (38) & Red Leak \\
SDSS Photometry Flag Cuts & 201 (231) & 44 (49) & \nodata (\nodata) & \nodata \\
Image Check & 195 (225) & 41 (46) & 23 (29) & \nodata \\
\hline

\enddata

\tablecomments{The criteria for selecting flares are shown, separately
for the $ugriz$ sample, the 
$ug$-only sample and the spectroscopic sample (those stars 
with SDSS spectra).
The number of stars is given, and the number of epochs is shown in 
parentheses.  The final flare sample contains 236 stars with 271 flaring
epochs.}

\label{table:steps_ugriz}
\end{deluxetable}

\begin{deluxetable}{cccccc}
 \tabletypesize{\scriptsize}
 \tablecaption{APO 3.5m DIS (R300) Spectral Sample}

\tablehead{
\colhead{SDSS ID}&
\colhead{Spectral Type}&
\colhead{H$\alpha$  EW (\AA)}&
\colhead{Notes}&
\colhead{Activity}&
\colhead{Dates}
}
\startdata
J203940.38+001326.7 & M1 & $1.4$ & -- & y & 080825 \\
J204133.37-001958.5 & M5 & $4.7$ & -- & y &  080919 \\
J204144.10+004048.2 & M4, M4 & $0.4$, $1.5$ & -- & n, n & 081107,  080919 \\
J204452.77+010546.9 & M1, M2 & $0.3$, $0.9$ & -- & var & 081107,  080919 \\
J204725.80+004128.0 & M2 & $3.3$ & -- & y &  080919 \\
J204728.12+002206.6 & M4 & $8.9$ & -- & y & 080919 \\
J205020.88-010839.9 & M3 & $3.5$ & -- & y &  080919\\
J205300.05-003821.8 & M3 & $3.8$ & -- & y &  080919\\
J205632.51+001413.5 & M1 & $2.2$ & -- & y & 080919\\
J210012.57-005326.0 & M5 & $3.8$ & -- & y &  080919\\
J210023.79-005830.1 & M0 & $0.8$ & -- & w &  080919\\
J210034.02-005814.6 & M6 & $6.3$ & -- & y &  080919\\
J210343.88+000303.1 & M3 & $7.0$ & -- & y &  080919\\
J211110.85+005027.1 & M4 & $4.5$ & -- & y &  080919\\
J211138.04+002137.2 & M4 & $5.9$ & -- & y &   080919\\
J211311.65-002503.2 & M4 & $4.1$ & -- & y &  080919\\
J211414.08-005534.7 & M5 & $5.0$ & -- & y &  080919\\
J211454.52-000215.3 & M4 & $5.3$ & -- & y & 081107 \\
J211827.00+000248.5 & M2, M0, M1  & $1.6$, $1.4, 0.8$ & -- & y, y, w & 081107, 080825, 081214 \\
J212133.89-011137.4 & M4 & $5.5$ & -- & y &  080919\\
J212749.79+002607.2 & M3 & $2.9$ & -- & y & 081107 \\
J213112.08-002154.0 & M4 & $4.5$ & -- & y & 081107 \\
J213206.27+004026.9 & M5 & $7.8$ & -- & y &  080919\\
J213517.94-005725.1 & M1 & $1.6$ & -- & y &  080919\\
J214029.24+003320.6 & M3 & $1.0$ & --  & w &  080919\\
J214124.51-002220.5 & M4 & $2.0$ & -- & y & 081214 \\
J214226.15+010542.3 & M1 & $0.5$ & -- & w &  080919\\
J214245.57-005436.7 & M1 & $1.2$ & -- & y &  080919\\
J214751.69-003237.4 & M4 & $3.9$ & -- & y &  080919\\
J215430.88+002117.0 & M4 & $3.3$ & -- & y &  080919\\
J220548.89-000352.5 & M1, M1 & $0.0$, $0.2$ & -- & n, n & 080825,  080919\\
J220920.29-005738.8 & M2 & 2.4 & -- & y & 081214 \\
J221726.10+003052.2 & M3 & $2.9$ & -- & y & 081013 \\
J222012.89-003429.0 & M1 & $1.3$ & -- & y & 080825\\
J222552.54+000251.6 & M4 & $4.1$ & -- & y &  080919\\
J223711.01+005306.5 & M2 & $2.6$ & -- & y &  080919\\
J223731.35+005425.6 & M5 & $3.9$ & -- & y & 081013 \\
J225641.87+011050.9 & M5 & $2.8$ & -- & y & 080825\\
J230001.05+010454.1 & M5 & $5.6$ & -- & y & 081107 \\
J230314.24-003110.7 & M2 & $2.7$ & -- & y & 080825\\
J231109.45+002515.4 & M3 & $4.2$ & -- & y & 081107 \\
J231810.79-004512.3 & M2 & $3.4$ & -- & y & 080825\\
J232516.19-001600.9 & M2 & $1.8$ & -- & y & 080825\\
J233810.08-004450.3 & M4 & $4.3$ & -- & y & 080825\\
J233952.93-001005.3 & M5 & $5.8$ & -- & y & 080825\\
J234612.94+002507.3 & M1 & $0.0$ & -- & n & 081107 \\
J004145.91-002832.7 & M5 & $6.0$ & -- & y & 080825\\
J005847.77+011015.5 & M4 & $4.5$ & -- & y & 080825\\
J010052.58+010233.1 & M0 & $-0.2$ & -- & n & 081107 \\
J011745.28+005929.4 & M1 & $1.8$ & -- & y & 081013\\
J012531.29-005359.0 & M3 & $2.7$ & -- & y & 080825\\
J013851.46-001621.7 & M6 & $12.0$ & -- & y & 081013 \\
J020056.76+004149.1 & M4 & $9.2$ & -- & y & 081013 \\
J020327.92-000442.1 & M3 & $3.1$ & -- & y & 081013 \\
J022112.48+003225.7 & M4,M3 & $0.7$,$0.8$ & -- & n,n & 080825, 080919\\
J022214.35-003217.5 & M2, M2 & $0.7, 0.6$ & -- & w, w & 081013, 081124 \\
J023452.37-010031.4 & M3 & $3.3$ & -- & y & 081107 \\
J023538.57-000422.2 & M4 & $6.0$ & -- & y & 081107 \\
J023547.98-005822.4 & M3 & $2.1$ & -- & y & 081107 \\
J023550.72+010136.8 & M3 & $2.3$ & -- & y & 081124 \\
J024554.14-010445.1 & M3 & $6.0$ & -- & y & 081124 \\
J025747.56+004731.2 & M4 & $3.8$ & -- & y & 081124 \\
J032605.67-002402.6 & M0 & $2.0$ & -- & y & 081124 \\
J033017.55+005359.4 & K7 & $-0.5$ & -- & n & 081124 \\
J033734.14-011104.5 & M2, M3 & $-0.6$, $0.0$ & -- &  n, n & 081013, 081124 \\
J033735.25+005859.7 & M2 & $2.4$ & -- & y & 081124 \\
J205059.98+005136.4 & M4 & $2.4$ & $ug$-only & y &  080919 \\
J205257.83-000335.9 & M3 & $3.3$ & $ug$-only & y &  080919 \\
J215455.18+011414.5 & M0 & 6.0 & $ug$-only & y & 081214 \\
J215517.42-004547.9 & M5 & 11.6 & $ug$-only & y & 081214 \\
J221712.10-000815.4 & M0 & -0.4 & $ug$-only & n & 081214 \\
J001208.37+002559.1 & M4 & 5.7 & $ug$-only & y & 081214 \\
J005903.60-011331.1 & M2 & 2.6 & $ug$-only & y & 081214 \\
J011307.16-002326.1 & M3 & 3.4 & $ug$-only & y & 081214 \\
J012904.68+003058.4 & M5 & 2.3 & $ug$-only & y & 081214 \\
J020911.90+004432.4 & M1 & 1.2 & $ug$-only & y & 081214\\

\enddata
\label{table:DIS}
\tablecomments{y - active; w - weakly active; n - inactive;  var - variable.
Activity status was determined using the criteria in \cite{West2004, West2008} and types were determined automatically with the Hammer facility (Covey et al. 2007).  The automatically assigned activity status and type were adjusted
by eye as needed, and EW's were
recalculated using accounting for the radial velocity shift. Errors
in the EW are are typically $\sim $0.2 \AA, and were calculated using Equation A10 in \cite{Chalabaev}.}
 
\end{deluxetable}
\clearpage

\clearpage

\begin{deluxetable}{ccc}
\tablecolumns{3}
\tabletypesize{\scriptsize}
\tablewidth{0pc}
\tablecaption{Spatial Flare Rates}
\tablehead{
\colhead{Spatial Division}    &  \colhead{Mean \# flares hr$^{-1}$ sq deg$^{-1}$}   & \colhead{Flaring Fraction}}

\startdata






$23\degr < |b| < 45\degr$ & 2.0 & 1.38$\times$10$^{-4}$  \\
$45\degr < |b| < 55\degr$ & 0.9 & 9.9$\times10^{-5}$ \\ 
$55\degr < |b| < 64\degr$ & 0.7 & 8.0$\times10^{-5}$   \\
$45\degr < l < 90\degr$ & 1.5 & 1.11$\times10^{-4}$  \\
$90\degr < l < 135\degr$ & 0.7 &  8.3$\times10^{-5}$  \\
$135\degr < l < 191\degr$ & 0.9 &  1.16$\times10^{-4}$ \\

\enddata
\label{table:spatialdivisiontable}
\tablecomments{ We present the raw spatial rate (mean number of flares hr$^{-1}$ sq deg$^{-1}$) and
 the flaring fraction for
different ranges of Galactic latitude and longitude. When calculating the raw spatial rate, we only consider the 1 $\times$ 1 sq deg
bins with at least 50\% SDSS coverage. }

\end{deluxetable}
\clearpage

\clearpage

\begin{deluxetable}{ccccccccc}
\tablecolumns{9}
 \tabletypesize{\scriptsize}
\tablewidth{0pc}
\tablecaption{Flare Rate Model Parameters}
\tablehead{
\colhead{Spectral Subtype Bin}    &  \colhead{$s$}   & \colhead{$f_{active}$} & \colhead{$f_{a,flare}$}  & \colhead{$f_{inactive}$} & \colhead{$\nu$} & \colhead{limiting log L$_U$} & \colhead{Predicted \# flares*} & \colhead{Observed \# flares}}

\startdata



M0-M1 & 35,000 & 0.03 & 1.0 & 0.97 & 0.02 & $29.5$ & 20-66 & 38 \\
M2-M3 & 13,000 & 0.06 & 1.0 & 0.94 & 0.1 & $28.7$ & 68-150 & 83 \\
M4-M6 & 2,100 & 0.30 & 1.0 & 0.70 & 0.4 & $28.0$ & 193-233 & 150 \\
\enddata
\label{table:spatialtable}
\tablecomments{We present the measured and derived values for the flare rate model given by Equation 4 in \S 4.6.1.}
\tablenotetext{*}{We adopt
1\% and 10\% for $f_{i, flare}$, the fraction of inactive stars that are capable of flaring, which give a range
of values for the predicted number of flares. }

\end{deluxetable}
\clearpage

\bibliography{june10rev}

\end{document}